\documentclass[%
 reprint,
superscriptaddress,
amsmath,amssymb,
aps,
pra,
]{revtex4-1}
\usepackage{caption}
\usepackage{subcaption}
\usepackage{graphicx}
\usepackage{xcolor}
\usepackage{graphicx}
\usepackage{dcolumn}
\usepackage{bm}

\begin{document}

\preprint{APS/123-QED}

\title{Refraction of space-time wave packets: I. Theoretical principles}

\author{Murat Yessenov}
\affiliation{CREOL, The College of Optics \& Photonics, University of Central Florida, Orlando, FL 32816, USA}
\author{Basanta Bhaduri}
\affiliation{CREOL, The College of Optics \& Photonics, University of Central Florida, Orlando, FL 32816, USA}
\author{Ayman F. Abouraddy}
\thanks{corresponding author: raddy@creol.ucf.edu}
\affiliation{CREOL, The College of Optics \& Photonics, University of Central Florida, Orlando, FL 32816, USA}




\begin{abstract}
Space-time (ST) wave packets are pulsed optical beams endowed with precise spatio-temporal structure by virtue of which they exhibit unique and useful characteristics, such as propagation invariance and tunable group velocity. We study in detail here, and in two accompanying papers, the refraction of ST wave packets at planar interfaces between non-dispersive, homogeneous, isotropic dielectrics. We formulate a law of refraction that determines the change in the ST wave-packet group velocity across such an interface as a consequence of a newly identified optical refractive invariant that we call `the spectral curvature'. Because the spectral curvature vanishes in conventional optical fields where the spatial and temporal degrees of freedom are separable, these phenomena have not been observed to date. We derive the laws of refraction for baseband, X-wave, and sideband ST wave packets that reveal fascinating refractive phenomena, especially for the former class of wave packets. We predict theoretically, and confirm experimentally in the accompanying papers, refractive phenomena such as group-velocity invariance (ST wave packets whose group velocity does not change across the interface), anomalous refraction (group-velocity increase in higher-index media), group-velocity inversion (change in the sign of the group velocity upon refraction but not its magnitude), and the dependence of the group velocity of the refracted ST wave packet on the angle of incidence.
\end{abstract}

\maketitle

\section{Introduction}

`Space-time' (ST) wave packets are a class of pulsed optical beams whose spatio-temporal spectra are structured so as to impose a tight association between the spatial and temporal frequencies (or transverse wave-vector components and wavelengths, respectively) \cite{Donnelly93ProcRSLA,Longhi04OE,Saari04PRE,Kondakci16OE,Parker16OE,Kondakci17NP,Kondakci19OL,Hall21OL}. Engineering this spectral association can yield wave packets that travel rigidly in free space without diffraction or dispersion \cite{Reivelt03arxiv,Kiselev07OS,FigueroaBook8,Turunen10PO,FigueroaBook14} at an arbitrary group velocity \cite{Salo01JOA}, including Brittingham's luminal focus-wave mode (FWM) \cite{Brittingham83JAP}, in addition to wave packets whose group velocity takes on superluminal \cite{Lu92IEEEa,Saari97PRL,Zamboni02EPJD,Recami03IEEEJSTQE,Valtna07OC,Zamboni09PRA}, subluminal \cite{Zamboni08PRA}, or even negative values \cite{Zapata06OL}. Crucially, deviation of the group velocity for ST wave packets in free space from $c$ (the speed of light in vacuum) stems from their spatio-temporal field structure and \textit{not} from chromatic dispersion in the medium.

There has recently been a resurgence of theoretical studies of spatio-temporally structured fields \cite{Porras17OL,Efremidis17OL,Wong17ACSP1,Wong17ACSP2,SaintMarie17Optica,PorrasPRA18,Wong20AS,Kibler21PRL,Shen21PRR,Bejot21arxiv} and newly emerging experimental realizations \cite{Froula18NP,Shaltout19Science,Hancock19Optica,Jolly20OE,Chong20NP,Chen21arxiv}. We recently introduced a phase-only spatio-temporal spectral synthesis methodology \cite{Kondakci17NP,Yessenov19OPN} that affords precise preparation of ST wave packets. This strategy has facilitated observing substantial and unambiguous departures from conventional behaviors, including propagation invariance \cite{Kondakci17NP,Kondakci18PRL,Bhaduri18OE,Bhaduri19OL,Yessenov19OE}; arbitrary group velocities in free space \cite{Kondakci19NC,Yessenov19OE,Yessenov20NC}, dielectrics \cite{Bhaduri19Optica,Bhaduri20NP}, planar waveguides \cite{Shiri20NC}, and as surface plasmon polaritons at metal-dielectric interfaces \cite{Schepler20ACSP}; self-healing \cite{Kondakci18OL}; time diffraction \cite{Porras17OL,PorrasPRA18,Kondakci18PRL,Yessenov20PRL1}; axial acceleration and deceleration \cite{Yessenov20PRL2}; and arbitrary dispersion in free space \cite{Yessenov21arxiv}.

Exploring the interaction of ST wave packets with photonic devices necessitates understanding their refraction at planar interfaces. Following the classification scheme of ST wave packets in Ref.~\cite{Yessenov19PRA}, we distinguish between three families: baseband, X-waves, and sideband (e.g., FWMs) -- which are expected to exhibit distinct refractive behavior by virtue of their spatio-temporal structure. To date, there have been only a few \textit{theoretical} studies of the refraction of X-waves \cite{Shaarawi00JASA,Attiya01PER,Shaarawi01PER,Salem12JOSAA} and FWMs \cite{Hillion93Optik,Donnelly97IEEE,Hillion98JO,Hillion99JOA}, which focused on modifications occurring in their reflected or refracted spatio-temporal profiles. However, no new refractive phenomena unique to these ST wave packets were unveiled, and \textit{no experimental investigations} were reported.

Using our versatile synthesis strategy, we recently reported an initial experimental study of the refraction of \textit{baseband} ST wave packets at a planar interface between two non-dispersive, homogeneous, isotropic dielectrics \cite{Bhaduri20NP}. By identifying a new optical invariant quantity (which we denote the `spectral curvature') that is unchanged upon traversing planar interfaces between pairs of such media, we uncovered a new law governing the change in the group velocity of refracted ST wave packets. In turn, this helps unveil new refractive phenomena exhibited by baseband ST wave packets -- even at normal incidence. First, for any two dielectrics regardless of their index contrast, there always exists a ST wave packet whose group velocity is \textit{invariant} across the interface. Second, there exists a regime of \textit{anomalous refraction} whereby the group velocity of the ST wave packet increases when traversing an interface from low to high index. Third, there always exists a ST wave packet whose group-velocity \textit{magnitude} does not change, but whose \textit{sign} switches after traversing the interface, leading to \textit{group-delay cancellation} when this ST wave packet traverses two equal-thickness layers of these materials. Finally, at oblique incidence, the group velocity of the transmitted ST wave packet changes with the angle of incidence at the interface \cite{Bhaduri20NP}. These surprising features highlight some of the rich physics underpinning the refraction of ST wave packets. 

\subsection*{Outline of this paper sequence}

\begin{figure}[t!]
  \begin{center}
  \includegraphics[width=5.4cm]{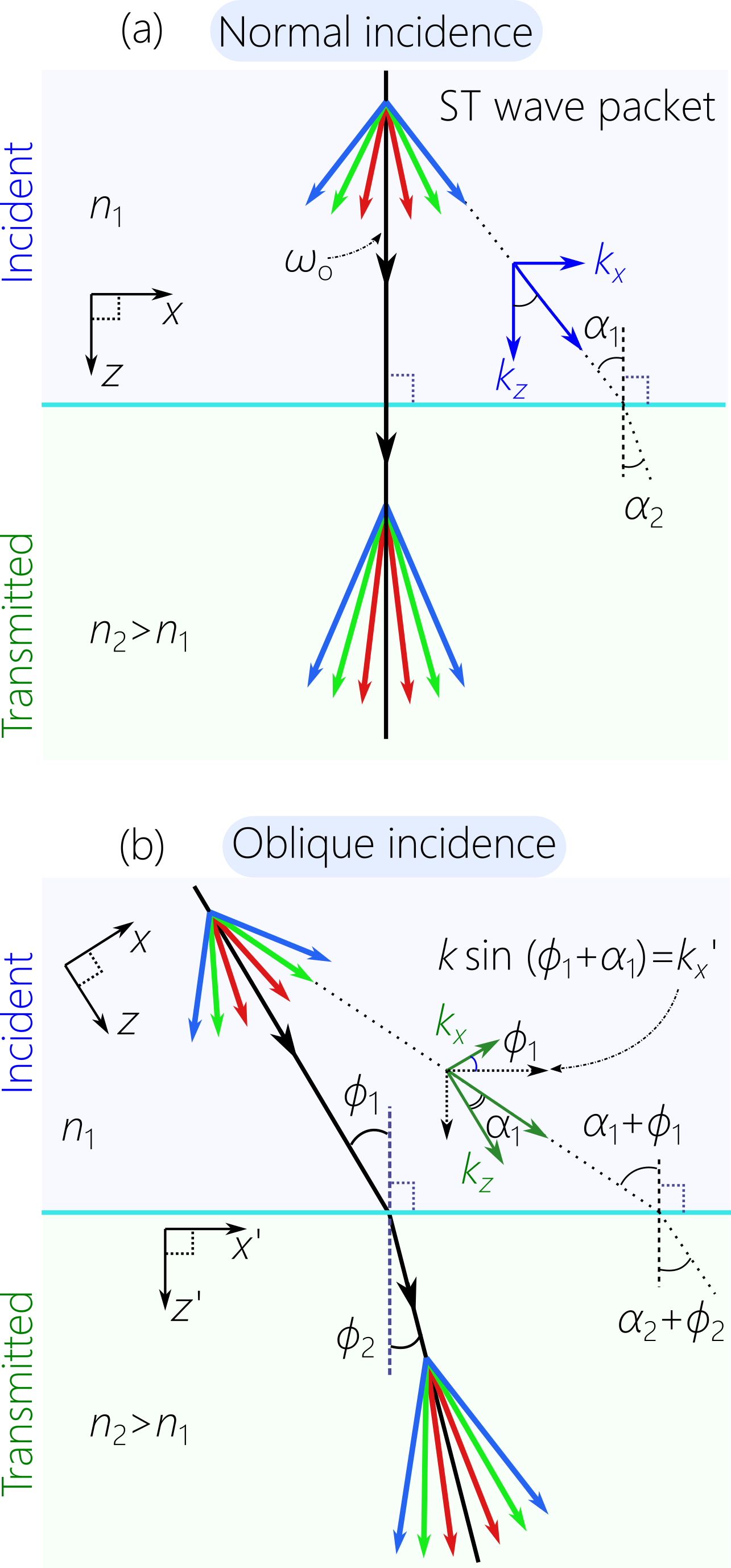}
  \end{center}
  \caption{Schematic depiction of the refraction of ST wave packets at a planar interface between two dielectrics of refractive indices $n_{1}$ and $n_{2}$ for (a) normal incidence and (b) oblique incidence at an angle $\phi_{1}$ with respect to the normal to the interface. In (b), the coordinate system $(x,z)$ is aligned with the direction of propagation of the ST wave packet, whereas $(x',z')$ is aligned with the interface. These two coordinate systems coincide in (a).}
  \label{Fig:NormalAndOblique}
\end{figure}

In this paper, and in two accompanying papers, we present a detailed theoretical and experimental study of the refraction of ST wave packets at a planar interface between two non-dispersive, homogeneous, isotropic dielectrics [Fig.~\ref{Fig:NormalAndOblique}]. This paper sequence provides a comprehensive study that extends beyond that in \cite{Bhaduri20NP}, which was restricted to an initial investigation of only baseband ST wave packets. In contrast, we examine here all families of ST wave packets, including FWMs and X-waves (however, when unspecified, ST wave packets will refer to the baseband class). Hereon, we refer to these papers with Roman numerals (I), (II), and (III).  

We first distinguish between \textit{intrinsic} and \textit{extrinsic} degrees of freedom (DoFs) in the optical field. Examples of extrinsic DoFs of a wave packet are its central wavelength, spatial and temporal bandwidths, direction of propagation, and spatio-temporal profile. Intrinsic DoFs, on the other hand, are related to the internal geometry of the field itself and are in principle independent of these extrinsic DoFs. For ST wave packets, we identify the `spectral tilt angle' as an intrinsic DoF of relevance to their refraction \cite{Yessenov19OPN}. This quantity is related to the curvature of its spatio-temporal spectrum, which determines the wave packet group velocity. In principle, the spectral tilt angle is independent of all the external DoFs.

Whereas Snell's law governs the change in an external DoF (the direction of propagation), the laws of refraction we examine in this paper sequence concern the change in an internal DoF; namely, the spectral tilt angle $\theta$. It is crucial to note here that the deviation of $\widetilde{v}$ in a non-dispersive medium from $c/n$ is dictated by the spectral tilt angle, $\widetilde{v}\!=\!c\tan{\theta}$. We aim to formulate a relationship between the group velocity $\widetilde{v}_{1}$ (or group index $\widetilde{n}_{1}\!=\!c/\widetilde{v}_{1}$) in the first medium and the corresponding quantities ($\widetilde{v}_{2}$ and $\widetilde{n}_{2}\!=\!c/\widetilde{v}_{2}$) in the second upon normal or oblique incidence at a planar interface [Fig.~\ref{Fig:NormalAndOblique}]. 

Snell's law stems from the invariance of the transverse component of the wave vector and the optical frequency across the interface. These conservation laws allow us to identify a quantity characteristic of ST wave packets that is invariant across planar interfaces, which we denote the `spectral curvature'. Our approach is closest to that in Ref.~\cite{Donnelly97IEEE}, which in turn is based on the pioneering work on the spatio-temporal spectral representations of ST wave packets on the light-cone \cite{Donnelly93ProcRSLA}. As mentioned above, previous theoretical studies of the refraction of ST wave packets focused on the changes in the spatio-temporal profiles (an extrinsic DoF). The spectral curvature and the spectral tilt angle are intrinsic DoFs that are \textit{independent} of the wave packet profile. Indeed, although the refracted wave packet profile may undergo changes (e.g., due to the Fresnel coefficients), these do \textit{not} affect our conclusions.

The content of the papers in this sequence can be summarized as follows.

\subsubsection*{Paper~(I)}

We lay out in paper (I) the theoretical foundations for the experimental work that will be reported in (II) and (III). First, we review of the formulation of ST wave packets in a dielectric and their spectral representation on the surface of the light-cone, which is a useful tool to help visualize the spatio-temporal spectral dynamics upon refraction. We next derive the laws of refraction at normal incidence for all three families of ST wave packets: baseband, sideband (FWMs), and X-waves \cite{Yessenov19PRA}. More emphasis is placed on the first because they exhibit the most interesting refractive phenomena and because of the difficulty in synthesizing X-waves and FWMs \cite{Yessenov19PRA}. We then formulate a law of refraction at oblique incidence for baseband ST wave packets; a similar law can\textit{not} be devised for X-waves or sideband ST wave packets.

\subsubsection*{Paper~(II)}

In paper (II) we confirm experimentally the novel refractive phenomena associated with baseband ST wave packets at \textit{normal} incidence [Fig.~\ref{Fig:NormalAndOblique}(a)] using a variety of optical materials with refractive indices in the range $1.38-1.76$, including MgF$_2$, UV fused silica, BK7 glass, and sapphire. We report observations of the novel refractive phenomena predicted in paper (I): (1) group-velocity invariance; (2) the transition from normal to anomalous refraction; and (3) group-velocity inversion and group-delay cancellation. We also provide a description of the dynamics of the spectral support domain of ST wave packets on the light-cone, which elucidates the counter-intuitive transition from normal to anomalous refraction.

\subsubsection*{Paper~(III)}

In paper (III) we examine the refraction of baseband ST wave packets at \textit{oblique} incidence [Fig.~\ref{Fig:NormalAndOblique}(b)]. We verify experimentally the modified law of refraction that accounts for oblique incidence and show how the group velocity of the transmitted ST wave packet changes with the angle of incidence while holding all other variables fixed. We then examine how this effect can be exploited in a scheme for blindly synchronizing multiple receivers at different \textit{a priori} unknown distances beyond an interface between different media. A first experimental test of this blind synchronization scheme is reported.

\begin{figure}[t!]
  \begin{center}
  \includegraphics[width=8.6cm]{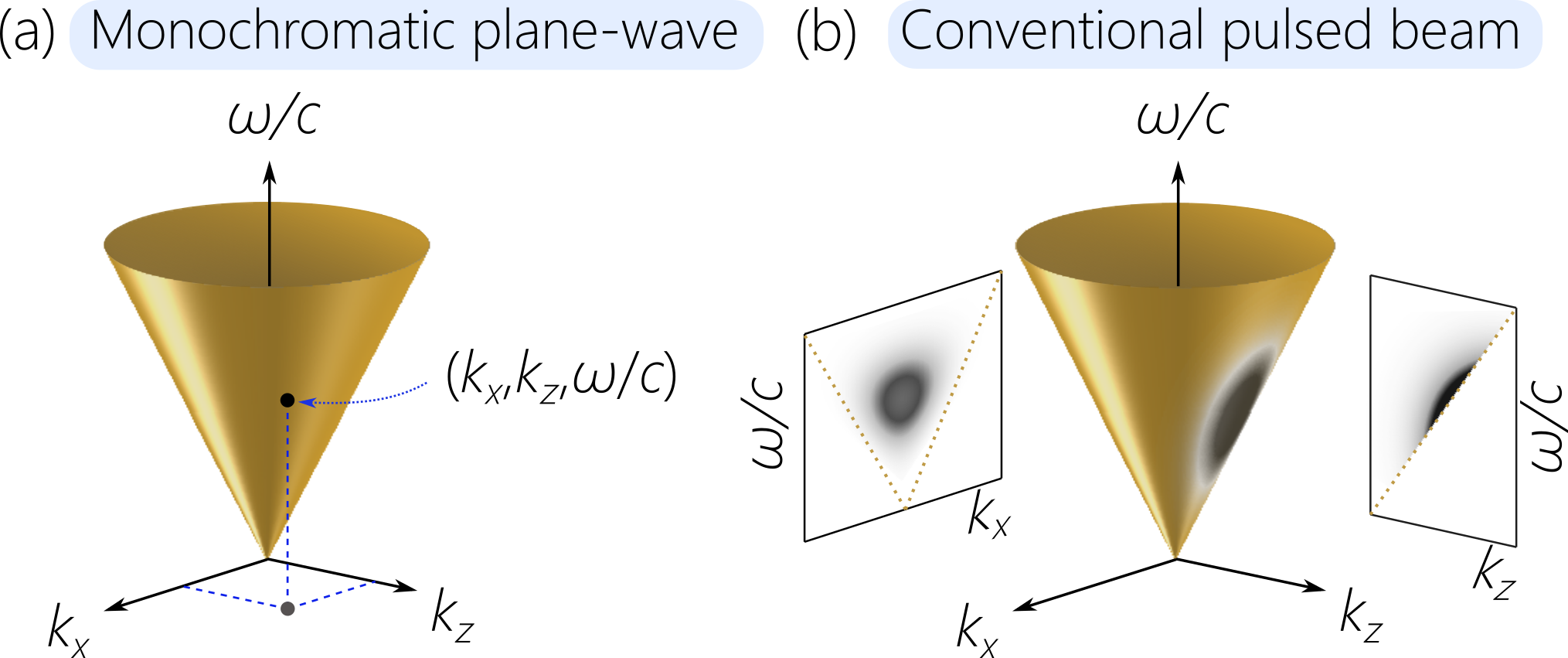}
  \end{center}
  \caption{(a) The light-cone $k_{x}^{2}+k_{z}^{2}\!=\!n^{2}(\tfrac{\omega}{c})^{2}$ in a non-dispersive medium of refractive index $n$ is depicted in the spectral space $(k_{x},k_{z},\tfrac{\omega}{c})$. A monochromatic plane-wave $e^{i(k_{x}x+k_{z}z-\omega t)}$ corresponds to a point on the light-cone surface. (b) The spectral support domain of a conventional pulsed beam, in which the spatial and temporal DoFs are separable, is a 2D region on the light-cone surface. We also show the spectral projections onto the $(k_{x},\tfrac{\omega}{c})$ and $(k_{z},\tfrac{\omega}{c})$ planes. The dotted lines are the light-lines in the medium.}
  \label{Fig:LightCone}
\end{figure}

\section{Space-time wave packets in a non-dispersive dielectric}

ST wave packets are distinguished from conventional pulsed beams by being endowed with a precise spatio-temporal structure, which is best elucidated in the spectral domain \cite{Donnelly93ProcRSLA,Saari04PRE,Kondakci16OE,Efremidis17OL}. We consider optical fields $E(x,z;t)$ described by one transverse coordinate $x$, the axial coordinate $z$, and time $t$ (the field is uniform along $y$, which simplifies the analysis and the experiments without loss of generality). The angular frequency $\omega$ is referred to as the \textit{temporal} frequency to symmetrize the nomenclature with respect to the \textit{spatial} frequency $k_{x}$, which is the transverse component of the wave vector (the axial component is $k_{z}$). In a non-dispersive, homogeneous, isotropic dielectric of refractive index $n$, the dispersion relationship is $k_{x}^{2}+k_{z}^{2}\!=\!n^{2}(\tfrac{\omega}{c})^{2}$, which corresponds to the surface of a light-cone of opening angle $\arctan{(n)}$. A monochromatic plane wave $e^{i(k_{x}x+k_{z}z-\omega t)}$ is represented by a point on the light-cone surface [Fig.~\ref{Fig:LightCone}(a)]. The envelope $\psi(x,z;t)$ of a pulsed field $E(x,z;t)\!=\!e^{i(nk_{\mathrm{o}}z-\omega_{\mathrm{o}}t)}\psi(x,z;t)$ can be decomposed into an angular spectrum \cite{SalehBook07}:
\begin{equation}
\psi(x,z;t)=\iint dk_{x}d\Omega\widetilde{\psi}(k_{x},\Omega)e^{i\{k_{x}x+(k_{z}-nk_{\mathrm{o}})z-\Omega t\}},
\end{equation}
where the spatio-temporal spectrum $\widetilde{\psi}(k_{x},\Omega)$ is the two-dimensional Fourier transform of $\psi(x,0;t)$, $\omega_{\mathrm{o}}$ is the central frequency, $k_{\mathrm{o}}\!=\!\tfrac{\omega_{\mathrm{o}}}{c}$ is the corresponding free-space wave number, and $\Omega\!=\!\omega-\omega_{\mathrm{o}}$. A conventional pulsed beam of spatial bandwidth $\Delta k_{x}$ and temporal bandwidth $\Delta\omega$ is represented by a two-dimensional (2D) domain on the surface of the light-cone [Fig.~\ref{Fig:LightCone}(b)]. In most conventional pulsed laser beams, the spatio-temporal spectrum is approximately separable with resepct to the spatial and temporal DoFs, $\widetilde{\psi}(k_{x},\Omega)\!\approx\!\widetilde{\psi}_{x}(k_{x})\widetilde{\psi}_{t}(\Omega)$.

\begin{figure*}[t!]
  \begin{center}
  \includegraphics[width=15.6cm]{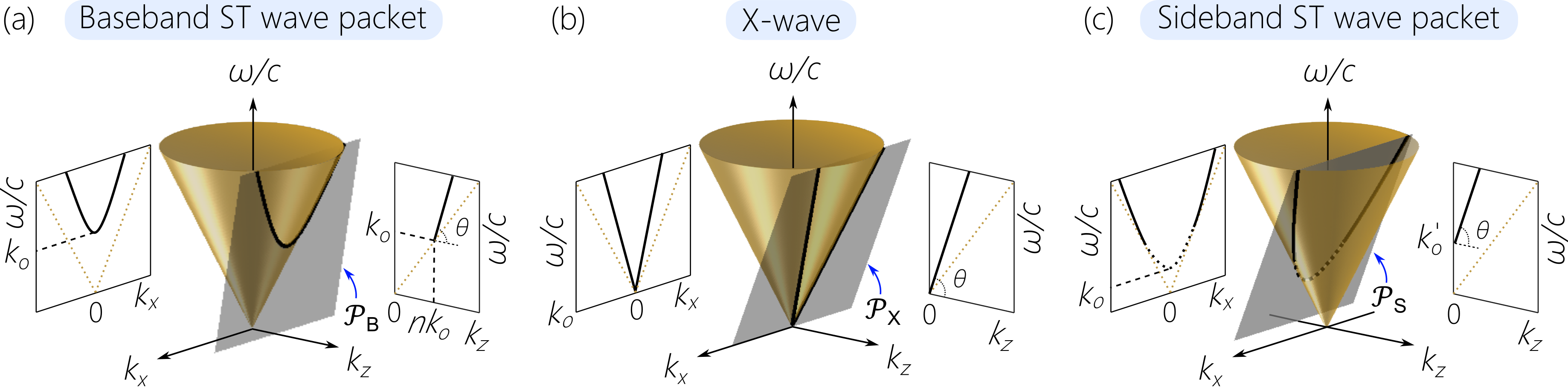}
  \end{center}
  \caption{Three classes of ST wave packets in a dielectric of index $n$, their spectral support domains on the light-cone surface, and their spectral projections (dark solid curves) on the $(k_{z},\tfrac{\omega}{c})$ and $(k_{x},\tfrac{\omega}{c})$ planes. (a) Baseband ST wave packets at the intersection of the light-cone with the spectral plane $\mathcal{P}_{\mathrm{B}}(\theta)$. (b) X-waves at the intersection of the light-cone with the spectral plane $\mathcal{P}_{\mathrm{X}}(\theta)$. (c) Sideband ST wave packets at the intersection between the light-cone with the spectral plane $\mathcal{P}_{\mathrm{S}}(\theta)$; $k_{\mathrm{o}}'\!=\!k_{\mathrm{o}}(1+n\tan{\theta})$. In all cases, only the portion of the spectral support domain where $k_{z}\!>\!0$ is physically permitted.}
  \label{Fig:ThreeClasses}
\end{figure*}

For ST wave packets, the spectral support domain of $\widetilde{\psi}(k_{x},\Omega)$ is no longer 2D. Rather, it is reduced to one dimension (1D) by enforcing an association between the spatial and temporal frequencies such that $\widetilde{\psi}(k_{x},\Omega)\!\rightarrow\!\widetilde{\psi}(k_{x})\delta(\Omega-\Omega(k_{x};\theta))$, where $\Omega(k_{x};\theta)$ is a deterministic mapping that dictates the functional dependence between $k_{x}$ and $\Omega$, and $\theta$ is a real continuous parameter identifying this association \cite{Donnelly93ProcRSLA,Longhi04OE,Saari04PRE,Kondakci17NP}. In other words, angular dispersion is introduced into the field \cite{Torres10AOP,Fulop10Review,Hall21OL,Yessenov21arxiv}. To achieve propagation invariance, $\Omega(k_{x};\theta)$ must impose the constraint $k_{z}\!=\!b+\Omega/\widetilde{v}$, where $b\!\leq\!nk_{\mathrm{o}}$ is a constant and $\widetilde{v}$ is the group velocity. In other words, the spectral support domain corresponds to the intersection of the light-cone with a plane $\mathcal{P}$ that is parallel to the $k_{x}$-axis. Propagation-invariant wave packets can be classified into three distinct families \cite{Yessenov19PRA}: baseband ST wave packets \cite{Kondakci16OE,Parker16OE,Kondakci17NP}, X-waves \cite{Saari97PRL}, and sideband ST wave packets \cite{Brittingham83JAP,Reivelt00JOSAA,Reivelt02PRE}. In all cases, the wave packet travels rigidly at a group velocity $\widetilde{v}\!=\!c\tan{\theta}$ and group index $\widetilde{n}\!=\!c/\widetilde{v}\!=\!\cot{\theta}$, where $\theta$ (the spectral tilt angle) is the angle $\mathcal{P}$ makes with respect to the $k_{z}$-axis \cite{Yessenov19PRA,Yessenov19OE,Yessenov19OPN}.

\subsection{Baseband ST wave packets}

\textit{Baseband} ST wave packets are so-called because temporal frequencies in the vicinity of the carrier frequency $\omega_{\mathrm{o}}$ are associated with spatial frequencies in the vicinity of $k_{x}\!=\!0$. This class of ST wave packets is identified with a plane $\mathcal{P}_{\mathrm{B}}(\theta)$ having the form $\Omega\!=\!(k_{z}-nk_{\mathrm{o}})c\tan{\theta}$, which passes through the point $(k_{x},k_{z},\tfrac{\omega}{c})\!=\!(0,nk_{\mathrm{o}},k_{\mathrm{o}})$ and intersects with the light-cone in the conic section [Fig.~\ref{Fig:ThreeClasses}(a)]:
\begin{equation}\label{Eq:ExactBaseband}
\left(\frac{\omega}{c}-\frac{\widetilde{n}}{\widetilde{n}+n}k_{\mathrm{o}}\right)^{2}+\frac{1}{\widetilde{n}^{2}-n^{2}}k_{x}^{2}=\frac{n^{2}}{(\widetilde{n}+n)^{2}}k_{\mathrm{o}}^{2}.
\end{equation}
As such, the envelope of the ST wave packet takes on the form
\begin{equation}\label{Eq:BasebandAngularSpectrum}
\psi_{\mathrm{B}}(x,z;t)\!=\!\int\! dk_{x}\widetilde{\psi}_{\mathrm{B}}(k_{x})e^{ik_{x}x}e^{-i\Omega(t-z/\widetilde{v})}=\!\psi_{\mathrm{B}}(x,0;t-z/\widetilde{v}),
\end{equation}
where $\widetilde{\psi}_{\mathrm{B}}(k_{x})$ is the Fourier transform of $\psi_{\mathrm{B}}(x,0;0)$. Subluminal propagation in a medium of index $n$ ($\widetilde{v}\!<\!\tfrac{c}{n}$, $\widetilde{n}\!>\!n$) is associated with the range $0\!<\!\tan{\theta}\!<\!\tfrac{1}{n}$, and superluminal propagation ($\widetilde{v}\!>\!\tfrac{c}{n}$, $\widetilde{n}\!<\!n$) with $\tan{\theta}\!>\!\tfrac{1}{n}$, which includes a negative-$\widetilde{v}$ regime ($\widetilde{v}\!<\!0$, $\widetilde{n}\!<\!0$) when $\theta\!>\!90^{\circ}$. The luminal condition ($\widetilde{v}\!=\!c/n$, $\widetilde{n}\!=\!n$) corresponds to $\mathcal{P}_{\mathrm{B}}(\theta)$ tangential to the light-cone, whereupon $\tan{\theta}\!=\!\tfrac{1}{n}$, and the conic section degenerates into a line $k_{z}\!=\!n\omega/c$ ($k_{x}\!=\!0$) representing a pulsed plane-wave. The continuous accessibility of group velocity values by tuning $\theta$ \cite{Kondakci17NP,Yessenov19OPN,Kondakci19NC} makes baseband ST wave packets most interesting with respect to refraction. Crucially, the group velocity of these wave packets is the speed of their peak \cite{Shaarawi00JPA,SaariPRA18,Saari19PRA,Saari20PRA}.

\subsection{X-waves}

The spectral support domain of X-waves comprises the two straight lines at the intersection of the light-cone with the plane $\mathcal{P}_{\mathrm{X}}(\theta)$ that passes through the origin $(k_{x},k_{z},\tfrac{\omega}{c})\!=\!(0,0,0)$, such that $\tfrac{\omega}{c}\!=\!k_{z}\tan{\theta}$,
\begin{equation}\label{Eq:ExactXWave}
k_{x}=\pm\sqrt{n^{2}-\widetilde{n}^{2}}\frac{\omega}{c};
\end{equation}
see Fig.~\ref{Fig:ThreeClasses}(b). Only positive-valued \textit{superluminal} group velocities ($\widetilde{v}\!>\!\tfrac{c}{n}$, $\widetilde{n}\!<\!n$) are allowed \cite{Yessenov19PRA}. The luminal condition corresponds to $\tan{\theta}\!=\!\tfrac{1}{n}$, similarly to baseband ST wave packets, whereupon the X-wave degenerates into a pulsed plane-wave.

\subsection{Sideband ST wave packets}

In sideband ST wave packets, the spatial frequencies in the vicinity of $k_{x}\!=\!0$ are physically \textit{forbidden} because they are associated with negative-valued $k_{z}$, which are incompatible with causal excitation and propagation \cite{Yessenov19PRA}. The spectral support domain here is the conic section at the intersection of the light-cone with a plane $\mathcal{P}_{\mathrm{S}}(\theta)$ given by $\Omega\!=\!(k_{z}+nk_{\mathrm{o}})c\tan{\theta}$ [Fig.~\ref{Fig:ThreeClasses}(c)]:
\begin{equation}\label{Eq:ExactSideband}
\left(\frac{\omega}{c}-\frac{\widetilde{n}}{\widetilde{n}-n}k_{\mathrm{o}}\right)^{2}-\frac{1}{\widetilde{n}^{2}-n^{2}}k_{x}^{2}=\frac{n^{2}}{(\widetilde{n}-n)^{2}}k_{\mathrm{o}}^{2}.
\end{equation}
This conic section is a hyperbola when $\widetilde{n}\!<\!n$ (superluminal), and an ellipse when $\widetilde{n}\!>\!n$ (subluminal). Although this ellipse can be shown to correspond to that of subluminal baseband ST wave packets after an appropriate transformation \cite{Yessenov19PRA}, we nevertheless distinguish here between the two. For subluminal \textit{baseband} ST wave packets, the spectral support domain is localized at $k_{x}\!=\!0$, whereas that for subluminal \textit{sideband} ST wave packets is localized away from $k_{x}\!=\!0$. Uniquely, the luminal condition $\widetilde{n}\!=\!n$ is \textit{not} a plane-wave pulse, but rather corresponds to Brittingham's FWM \cite{Brittingham83JAP} represented by a parabola, $\tfrac{\Omega}{c}=\tfrac{k_{x}^{2}}{4n^{2}k_{\mathrm{o}}}$. Increasing $\theta$ produces positive-valued superluminal sideband ST wave packets (or `focus-X-waves' \cite{Besieris98PIERS}). The selection of the spectral plane $\mathcal{P}_{\mathrm{S}}(\theta)$ results in a negative \textit{phase} velocity $E_{\mathrm{S}}(x,z;t)\!=\!e^{-i(nk_{\mathrm{o}}z+\omega_{\mathrm{o}})}\psi_{\mathrm{S}}(x,z;t)$ \cite{Belanger84JOSAA}, and the envelope $\psi_{\mathrm{S}}(x,z;t)$ is given by the same relationship in Eq.~\ref{Eq:BasebandAngularSpectrum}, except that the association between the spatial and temporal frequencies $\Omega(k_{x};\theta)$ is determined through Eq.~\ref{Eq:ExactSideband}, and $\widetilde{\psi}_{\mathrm{S}}(k_{x})$ is defined only over the range $k_{x}\!>\!n(1+\tfrac{n}{\widetilde{n}})k_{\mathrm{o}}$ whereupon $k_{z}\!>\!0$ and $\Omega\!>\!n\omega_{\mathrm{o}}/\widetilde{n}$ \cite{Yessenov19PRA}.

\section{Law of refraction for space-time wave packets}

Even at normal incidence, where the central wave vector (corresponding to $k_{x}\!=\!0$) is orthogonal to the interface, there is always a finite span of angles of incidence involved because the ST wave packet has a finite spatial bandwidth [Fig.~\ref{Fig:NormalAndOblique}(a)]. Relying on the conservation of energy \textit{and} transverse momentum across a planar interface ($\omega$ and $k_{x}$ are invariant, respectively), we identify in the following sections a spatio-temporal refractive invariant quantity that characterizes the spectral projection onto the $(k_{x},\tfrac{\omega}{c})$-plane. We make use of the relationships between $k_{x}$ and $\omega$ for baseband ST wave packets in Eq.~\ref{Eq:ExactBaseband}, X-waves in Eq.~\ref{Eq:ExactXWave}, and sideband ST wave packets in Eq.~\ref{Eq:ExactSideband} to derive this refractive invariant for each family. With the help of such invariants, we establish the relationship between the group velocity $\widetilde{v}_{1}$ (group index $\widetilde{n}_{1}$ or spectral tilt angle $\theta_{1}$) of the incident ST wave packet and the corresponding quantities ($\widetilde{v}_{2}$, $\widetilde{n}_{2}$, or $\theta_{2}$) in the second. Within this conception, the group velocity of the transmitted wave packet $\widetilde{v}_{2}$ depends not only on the refractive indices $n_{1}$ and $n_{2}$, but also on the group velocity of the incident wave packet $\widetilde{v}_{1}$. This is of course altogether different from the usual scenario for refraction in non-dispersive media where the group velocities are determined solely by the refractive indices ($\widetilde{n}_{1}\!=\!n_{1}$ and $\widetilde{n}_{2}\!=\!n_{2}$) and there is no influence from the group velocity of the incident wave packet on that of the transmitted.

It is convenient to plot the relationships we derive in the $(\theta_{1},\theta_{2})$ domain [Fig.~\ref{Fig:BasebandPlot}], where each family of ST wave packets is represented by a 1D curve that passes through the point $(\theta_{1}^{\mathrm{L}},\theta_{2}^{\mathrm{L}})$; here $\theta_{1}^{\mathrm{L}}$ is the spectral tilt angle for the luminal condition in the first medium $\cot{\theta_{1}^{\mathrm{L}}}\!=\!n_{1}$, and $\theta_{2}^{\mathrm{L}}$ in the second $\cot{\theta_{2}^{\mathrm{L}}}\!=\!n_{2}$. This condition delineates the superluminal/subluminal combinations in the two media [Fig.~\ref{Fig:BasebandPlot}(a)]. For example, any point in the upper-right quadrant of Fig.~\ref{Fig:BasebandPlot}(a) represents a wave packet that is superluminal in both media, whereas a point in the quadrant below it represents a wave packet that is superluminal in the first medium, but becomes subluminal after refraction in the second. The diagonal line $\theta_{1}\!=\!\theta_{2}$ divides the domain into \textit{normal} and \textit{anomalous} refraction regimes [Fig.~\ref{Fig:BasebandPlot}(b)]. Assuming that $n_{2}\!>\!n_{1}$, then the upper half corresponds to \textit{anomalous} refraction (i.e., $\widetilde{v}_{2}\!>\!\widetilde{v}_{1}$ despite $n_{2}\!>\!n_{1}$), and the lower half to \textit{normal} refraction (i.e., $\widetilde{v}_{2}\!<\!\widetilde{v}_{1}$ as expected when $n_{2}\!>\!n_{1}$). Conversely, when $n_{2}\!<\!n_{1}$, then the upper half of the domain above the diagonal corresponds to normal refraction and the lower half to anomalous refraction.

The point at the intersection of the curve representing the law of refraction with the diagonal $\theta_{1}\!=\!\theta_{2}$ corresponds to \textit{group-velocity invariance}: the group velocity of the incident wave packet  represented by this point is equal to that of the transmitted wave packet $\widetilde{v}_{2}\!=\!\widetilde{v}_{1}$, independently of the index contrast [Fig.~\ref{Fig:BasebandPlot}(b)]. Furthermore, the point at the intersection of the law-of-refraction curve with the anti-diagonal $\theta_{1}+\theta_{2}\!=\!180^{\circ}$ corresponds to \textit{group-velocity inversion}: the group velocity of the transmitted wave packet has the same \textit{magnitude} but opposite \textit{sign} of that of the incident wave packet $\widetilde{v}_{2}\!=\!-\widetilde{v}_{1}$ [Fig.~\ref{Fig:BasebandPlot}(b)]. Switching the direction of incidence (from $n_{2}$ to $n_{1}$) results in reflecting the curve representing the law of refraction (from $n_{1}$ to $n_{2}$) around the diagonal $\theta_{1}\!=\!\theta_{2}$. Note that the representation of refraction at normal incidence for a conventional collimated field corresponds to the point at $(\theta_{1},\theta_{2})\!=\!(\theta_{1}^{\mathrm{L}},\theta_{2}^{\mathrm{L}})$, and \textit{none} of the phenomena we proceed to describe therefore occur in this scenario.

\section{Normal incidence of baseband space-time wave packets}

\subsection{Law of refraction}

\begin{figure}[t!]
  \begin{center}
  \includegraphics[width=8.6cm]{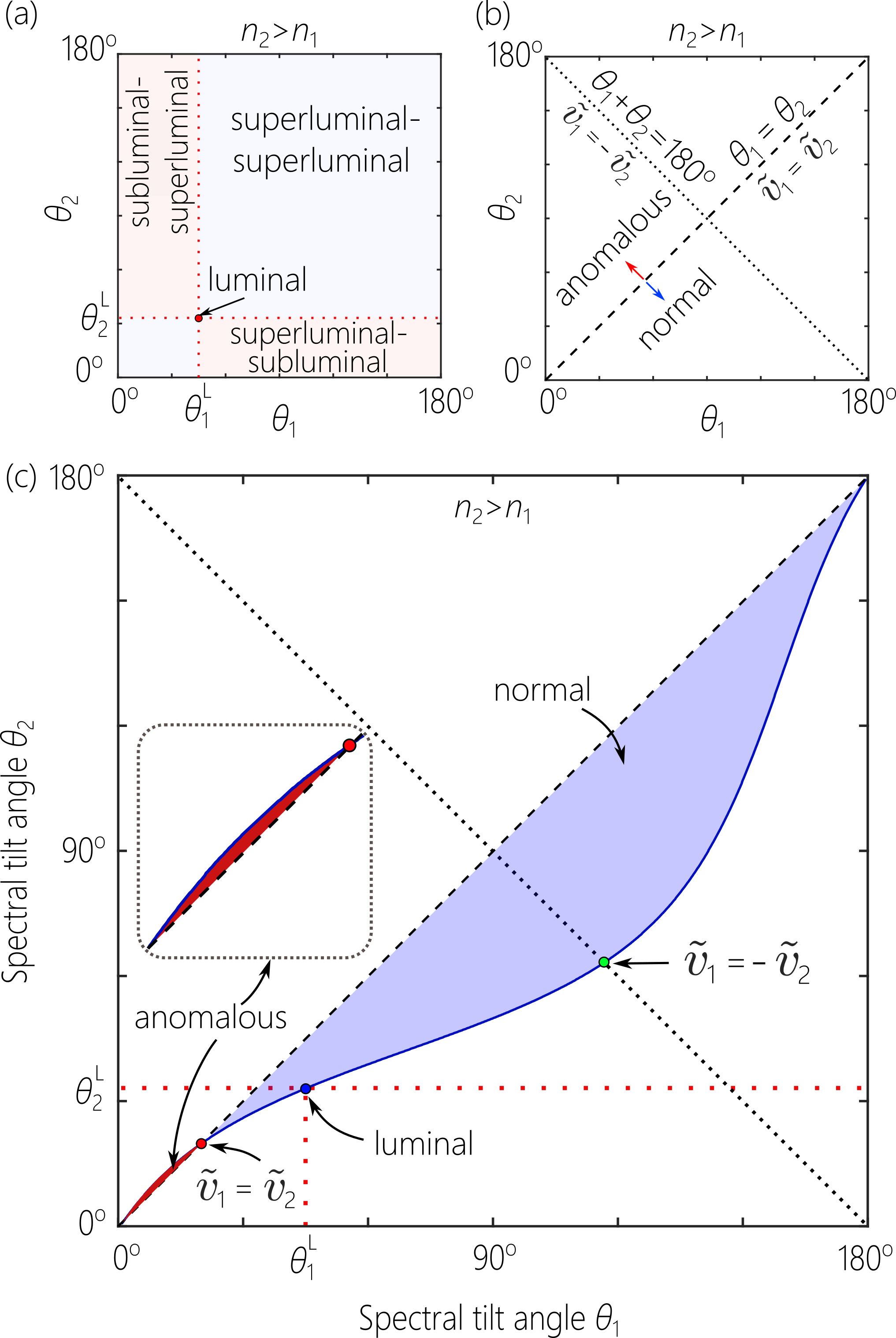}
  \end{center}
  \caption{(a) Definition of the subluminal, luminal, and superluminal regimes in the $(\theta_{1},\theta_{2})$ domain. (b) Identifying the conditions for normal/anomalous refraction; group-velocity invariance $\widetilde{v}_{1}\!=\!\widetilde{v}_{2}$; and group-velocity inversion $\widetilde{v}_{1}\!=\!-\widetilde{v}_{2}$ in the $(\theta_{1},\theta_{2})$ domain. (c) Law of refraction for \textit{baseband} ST wave packets according to Eq.~\ref{Eq:BasebandLawOfRefraction}. The inset highlights the anomalous refraction regime. We used $n_{1}\!=\!1$ and $n_{2}\!=\!1.5$ to produce this curve, but its main features are generic and apply to any pair of dielectrics.}
  \label{Fig:BasebandPlot}
\end{figure}

Assuming narrow temporal and spatial bandwidths, $\Delta\omega\!\ll\!\omega_{\mathrm{o}}$ and $\Delta k_{x}\!\ll\!k_{\mathrm{o}}$, respectively, the exact conic section for \textit{baseband} ST wave packets in Eq.~\ref{Eq:ExactBaseband} can be approximated in the vicinity of $k_{x}\!=\!0$ by a parabola,
\begin{equation}\label{Eq:BasebandParabola}
\frac{\Omega}{c}=\frac{1}{n(n-\widetilde{n})}\,\,\frac{k_{x}^{2}}{2k_{\mathrm{o}}}.
\end{equation}
The invariance of $k_{x}$ and $\Omega$ across the planar interface indicates that the quantity $n(n-\widetilde{n})$ is a new optical refractive invariant, which is related to the curvature of the parabolic spatio-temporal spectrum in Eq.~\ref{Eq:BasebandParabola}. We henceforth refer to this quantity as the `spectral curvature'. From this invariant, we formulate the law of refraction at normal incidence for baseband ST wave packets:
\begin{equation}\label{Eq:BasebandLawOfRefraction}
n_{1}(n_{1}-\widetilde{n}_{1})=n_{2}(n_{2}-\widetilde{n}_{2}). \end{equation}
This equation describes the change to the group index $\widetilde{n}_{1}\!\rightarrow\!\widetilde{n}_{2}$ (and thus the group velocity $\widetilde{v}_{1}\!\rightarrow\!\widetilde{v}_{2}$ and the spectral tilt angle $\theta_{1}\!\rightarrow\!\theta_{2}$) across the interface. The group index of the transmitted wave packet $\widetilde{n}_{2}$ depends on $n_{1}$, $n_{2}$, \textit{and} the group index of the incident wave packet $\widetilde{n}_{1}$. We plot in Fig.~\ref{Fig:BasebandPlot}(c) this law of refraction for $n_{1}\!=\!1$ and $n_{2}\!=\!1.5$, but the main features of the plotted curve are generic for all pairs of values for $n_{1}$ and $n_{2}$.

\subsection{Physical interpretation of the refractive invariant}

\begin{figure}[t!]
  \begin{center}
  \includegraphics[width=8.6cm]{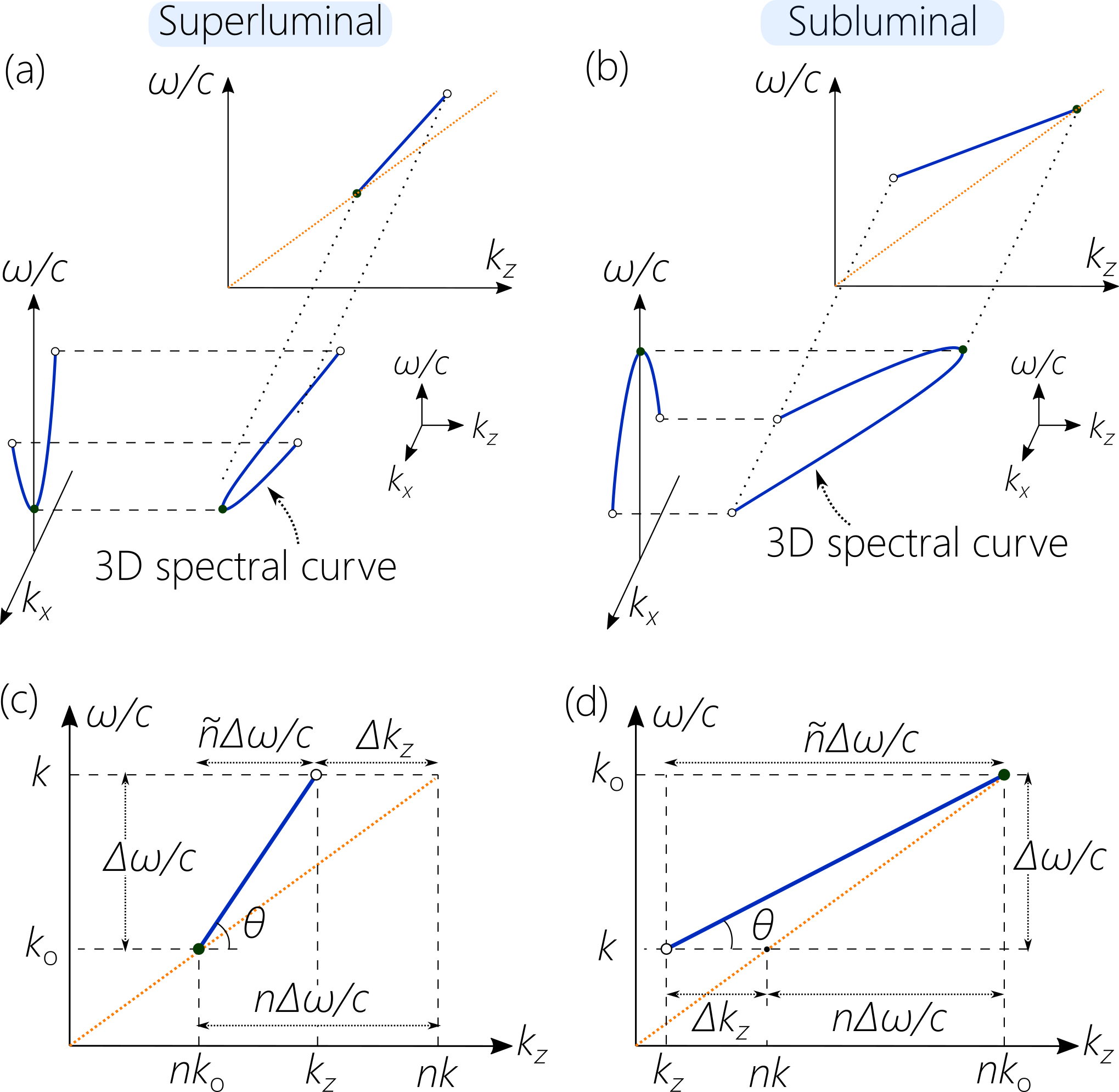}
  \end{center}
  \caption{Physical interpretation of the refractive invariant $n(n-\widetilde{n})$ for (a) superluminal and (b) subluminal baseband ST wave packets. We plot the three-dimensional spectral curve on the light-cone surface (the intersection of the spectral plane $\mathcal{P}_{\mathrm{B}}(\theta)$ with the light-cone) representing the spectral support domain of the ST wave packet, along with its spectral projections onto the $(k_{x},\tfrac{\omega}{c})$ and $(k_{z},\tfrac{\omega}{c})$ planes. The surface of the light-cone is removed for clarity [see Fig.~\ref{Fig:ThreeClasses}(a)]. (c) Definition of terms in the spectral support domain projected onto the $(k_{z},\tfrac{\omega}{c})$-plane for superluminal and (d) subluminal ST wave packets. In (c) we have $\Delta k_{z}\!=\!(n-\widetilde{n})\Delta\omega/c$, and in (d) $k_{z}\!=\!(\widetilde{n}-n)\Delta\omega/c$.}
  \label{Fig:PhysicalInterpretation}
\end{figure}

The physical interpretation of the quantity $n(n-\widetilde{n})$ can be elucidated by examining the spectral support domain projected onto the $(k_{z},\tfrac{\omega}{c})$-plane. Starting from the dispersion relationship $k_{x}^{2}+k_{z}^{2}\!=\!n^{2}(\tfrac{\omega}{c})^{2}$, the paraxial and narrow-bandwidth approximations yield:
\begin{equation}\label{Eq:ParaxialApproximation}
k_{z}\approx n\frac{\omega}{c}-\frac{k_{x}^{2}}{2nk_{\mathrm{o}}}.
\end{equation}
When $k_{x}\!=\!0$, the projected spectrum lies along the light-line $k_{z}\!=\!n\tfrac{\omega}{c}$, corresponding to a plane-wave pulse whose group velocity is $\widetilde{v}\!=\!c/n$. Introducing a finite spatial frequency $k_{x}$ into any plane-wave shifting the point representing its projection away from the light-line. The extent of this shift is determined by the magnitude of $k_{x}$ \textit{and} the refractive index $n$ as seen from the $k_{x}^{2}/(2nk_{\mathrm{o}})$ term in Eq.~\ref{Eq:ParaxialApproximation}. This is because the \textit{curvature} of the light-cone is determined by $n$, and this curvature dictates the change in $k_{z}$ in response to a change in $k_{x}$; see Fig.~\ref{Fig:PhysicalInterpretation}(a,b).

So far, we have not involved the unique constraint associated with baseband ST wave packets. Now consider two plane-waves, one at a temporal frequency $\omega_{\mathrm{o}}$ that lies on the light-line ($k_{x}\!=\!0$ and $k_{z}\!=\!n\tfrac{\omega_{\mathrm{o}}}{c}$), and another at $\omega\!>\!\omega_{\mathrm{o}}$ that does not ($k_{x}\!\neq\!0$ and $k_{z}\!\neq\!n\tfrac{\omega}{c}$); see Fig.~\ref{Fig:PhysicalInterpretation}(c). We define the deviation of the latter plane wave from the light-line as $\Delta k_{z}\!=\!nk-k_{z}$. If these two plane waves are the end points for the spectrum of a superluminal baseband ST wave packet, then the plane waves lying between $\omega_{\mathrm{o}}$ and $\omega$ are projected onto the straight light joining these two and tilted by an angle $\theta$ with respect to the $k_{z}$-axis. In this case $\Delta k_{z}\!=\!(n-\widetilde{n})\tfrac{\Delta\omega}{c}$, where $\widetilde{n}\!=\!\cot{\theta}$, the group velocity is $\widetilde{v}\!=\!c\tan{\theta}$, and $\Delta\omega\!=\!\omega-\omega_{\mathrm{o}}$ is the temporal bandwidth. Therefore, the term $(n-\widetilde{n})$ represents the deviation in wave-packet group velocity from the luminal limit. However, the quantity $n-\widetilde{n}\!=\!\tfrac{k_{x}^{2}}{2nk_{\mathrm{o}}\Delta\omega/c}$ is \textit{not} invariant with respect to changes in $n$. Nevertheless, because $k_{x}$ and $\omega$ are invariant across a planar interface, we identify instead the quantity $\tfrac{k_{x}^{2}}{2k_{\mathrm{o}}\Delta\omega/c}\!=\!n(n-\widetilde{n})$ as a refractive invariant. In other words, when traversing a planar interface between two media of refractive indices $n_{1}$ and $n_{2}$, the quantities $n_{1}(n_{1}-\widetilde{n}_{1})$ and $n_{2}(n_{2}-\widetilde{n}_{2})$ associated with the incident and transmitted wave packets are equal. Although the refractive index of the medium is different, $\widetilde{n}_{2}$ changes with respect to $\widetilde{n}_{1}$ to maintain this quantity fixed. This analysis assumed a superluminal ST wave packet ($\widetilde{n}\!<\!n$) [Fig.~\ref{Fig:PhysicalInterpretation}(a,c)], but a similar logic applies to the subluminal case ($\widetilde{n}\!>\!n$) [Fig.~\ref{Fig:PhysicalInterpretation}(b,d)].  

This new refractive invariant is a product of two terms: (1) the term $\widetilde{n}-n$ that arises from the constraint unique to ST wave packets, and represents the extent of the deviation away from the light-line (and thus the deviation of the group velocity $\widetilde{v}$ from $c$); and (2) the term $n$ arising from the curvature of the light-cone surface, which dictates the degree to which the spatial frequencies can produce a deviation from the light-line. This invariant quantity $n(\widetilde{n}-n)$ is the inverse of the curvature of the spectral projection of the ST wave packet onto the $(k_{x},\tfrac{\omega}{c})$-plane in the vicinity of $k_{x}\!=\!0$. For conventional pulsed beams in a non-dispersive medium, $\widetilde{n}\!\approx\!n$ and the spectral curvature therefore vanishes. Consequently, this refractive invariant quantity has \textit{not} had impact on optical refraction to date.

\subsection{Consequences of the law of refraction}

Despite the simplicity of the law of refraction in Eq.~\ref{Eq:BasebandLawOfRefraction}, it nevertheless reveals several surprising physical consequences.

\subsubsection{Luminal wave packets remain luminal}

If the incident wave packet is luminal $\widetilde{n}_{1}\!=\!n_{1}$ (i.e., a plane-wave pulse), the transmitted wave packet remains luminal in the second medium with $\widetilde{n}_{2}\!=\!n_{2}$ (i.e., remains a plane-wave pulse), as confirmed by direct substitution in Eq.~\ref{Eq:BasebandLawOfRefraction}. Furthermore, if the incident wave packet is subluminal $\widetilde{n}_{1}\!>\!n_{1}$ (superluminal $\widetilde{n}_{1}\!<\!n_{1}$), then it \textit{remains} subluminal $\widetilde{n}_{2}\!>\!n_{2}$ (superluminal $\widetilde{n}_{2}\!<\!n_{2}$). The subluminal-superluminal and superluminal-subluminal regimes in Fig.~\ref{Fig:BasebandPlot}(c) are forbidden. In other words, refraction of ST wave packets does not bridge the subluminal/superluminal barrier.

\subsubsection{Group-velocity invariance}

Can the group velocity of a ST wave packet remain unchanged after traversing the interface? Although this is impossible for conventional pulses, Eq.~\ref{Eq:BasebandLawOfRefraction} indicates that setting the group indices in the two media equal $\widetilde{n}_{1}\!=\!\widetilde{n}_{2}\!=\!\widetilde{n}_{\mathrm{th}}$ yields a solution:
\begin{equation}
\widetilde{n}_{\mathrm{th}}=n_{1}+n_{2}.
\end{equation}
That is, a normally incident ST wave packet with group index $\widetilde{n}_{1}\!=\!n_{1}+n_{2}$ will be transmitted \textit{without changing its group index} -- independently of the contrast between $n_{1}$ and $n_{2}$. Because $\widetilde{n}_{\mathrm{th}}\!>\!n_{1},n_{2}$, such a ST wave packet is subluminal in both media. This condition corresponds to the point at the intersection of the curve in $(\theta_{1},\theta_{2})$-space representing Eq.~\ref{Eq:BasebandLawOfRefraction} with the diagonal $\theta_{1}\!=\!\theta_{2}$ [Fig.~\ref{Fig:BasebandPlot}(b,c)].

\begin{figure}[t!]
  \begin{center}
  \includegraphics[width=8.6cm]{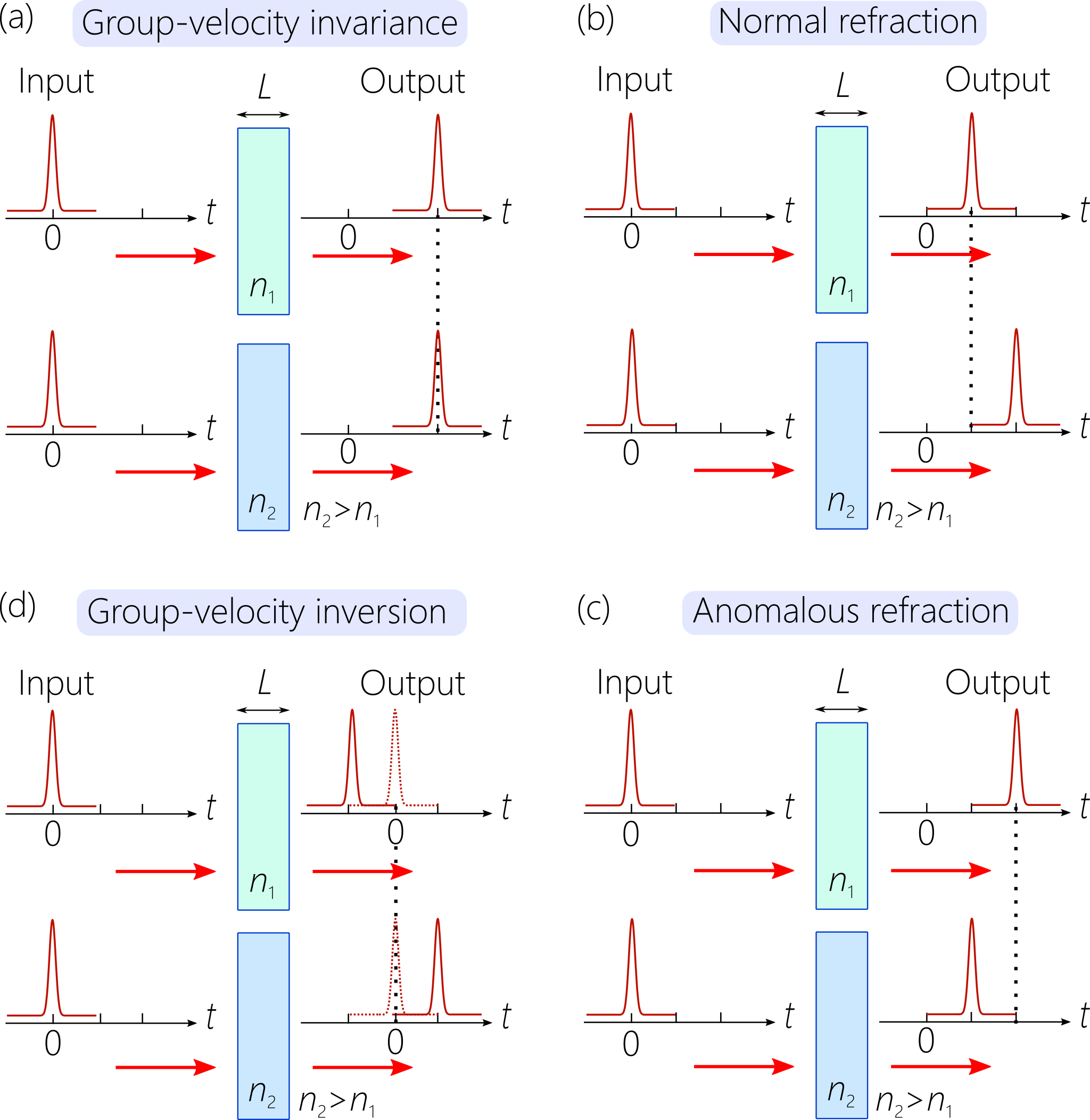}
  \end{center}
  \caption{Consequences of the law of refraction for baseband ST wave packets. Illustrations of (a) group-velocity invariance, (b) normal refraction, (c) anomalous refraction, and (d) group-velocity inversion for ST wave packets. The wave packet in each configuration traverses equal-thickness layers of indices $n_{1}$ and $n_{2}$ at normal incidence, with $n_{2}\!>\!n_{1}$.}
  \label{Fig:Consequences}
\end{figure}

We can view this phenomenon from a different perspective. If we assume incidence on either medium $n_{1}$ or $n_{2}$ from an ambient medium of index $n_{0}$, then we have $n_{0}(n_{0}-\widetilde{n}_{0})\!=\!n_{1}(n_{1}-\widetilde{n}_{1})$ and $n_{0}(n_{0}-\widetilde{n}_{0})\!=\!n_{2}(n_{2}-\widetilde{n}_{2})$, where $\widetilde{n}_{0}$ is the group index of the ST wave packet in the ambient medium. If $n_{0}\!=\!1$, then the threshold condition $\widetilde{n}_{\mathrm{th}}\!=\!n_{1}+n_{2}$ corresponds to $\widetilde{n}_{0}\!=\!1+n_{1}n_{2}$. Such a wave packet incident from free space on layers of indices $n_{1}$ and $n_{2}$ of equal thickness $L$ incurs the \textit{same} group delay $\tau_{1}\!=\!\tau_{2}\!=\!\widetilde{n}_{\mathrm{th}}L/c$; see Fig.~\ref{Fig:Consequences}(a).

\subsubsection{Anomalous refraction}

The group index $\widetilde{n}_{\mathrm{th}}\!=\!n_{1}+n_{2}$ is a threshold that separates regimes of normal and anomalous refraction. For $n_{2}\!>\!n_{1}$, substitution of $\widetilde{n}_{1}\!<\!\widetilde{n}_{\mathrm{th}}$ ($\theta_{1}\!>\!\theta_{\mathrm{th}}$, where $\widetilde{n}_{\mathrm{th}}\!=\!\cot{\theta_{\mathrm{th}}}$) for the incident ST wave packet in Eq.~\ref{Eq:BasebandLawOfRefraction} shows that the group velocity of the transmitted wave packet is \textit{lower} than that of the incident wave packet, as expected when $n_{2}\!>\!n_{1}$. Indeed, the corresponding segment of the curve representing the law of refraction for $\theta_{1}\!>\!\theta_{\mathrm{th}}$ is below the diagonal $\theta_{1}\!=\!\theta_{2}$ in Fig.~\ref{Fig:BasebandPlot}(c), thereby indicating \textit{normal} refraction [Fig.~\ref{Fig:Consequences}(b)]. That is, the group velocity drops in the higher-index medium.

On the other hand, when the incident ST wave packet is such that $\widetilde{n}_{1}\!>\!\widetilde{n}_{\mathrm{th}}$ ($\theta_{1}\!<\!\theta_{\mathrm{th}}$), the opposite behavior emerges: the group velocity anomalously \textit{increases} in the higher-index medium. The corresponding segment of the curve representing the law of refraction for $\theta_{1}\!<\!\theta_{\mathrm{th}}$ is above the diagonal $\theta_{1}\!=\!\theta_{2}$ in Fig.~\ref{Fig:BasebandPlot}(c), thereby indicating \textit{anomalous} refraction [Fig.~\ref{Fig:Consequences}(c)]. That is, the group velocity \textit{increases} in the higher-index medium. To the best of our knowledge, this type of behavior has never been identified or demonstrated in non-dipsersive dielectrics.

\subsubsection{Group-velocity inversion and group-delay cancellation}

In the superluminal regime $\widetilde{n}_{1}\!<\!n_{1}$, refraction of ST wave packets is always normal. However, because baseband ST wave packets admit both positive- and negative-valued group velocities, the curve representing Eq.~\ref{Eq:BasebandLawOfRefraction} in the $(\theta_{1},\theta_{2})$ domain in Fig.~\ref{Fig:BasebandPlot}(c) intersects with the anti-diagonal $\theta_{1}+\theta_{2}\!=\!180^{\circ}$. At this intersection point $\widetilde{n}_{2}\!=\!-\widetilde{n}_{1}$; i.e., the \textit{magnitude} of the group index of the incident wave packet remains constant upon transmission, but its \textit{sign} is flipped. This condition occurs when
\begin{equation}
\widetilde{n}_{1}=n_{1}-n_{2},
\end{equation}
and direct substitution into Eq.~\ref{Eq:BasebandLawOfRefraction} yields $\widetilde{n}_{2}\!=\!n_{2}-n_{1}\!=\!-\widetilde{n}_{1}$. We refer to this condition as \textit{group-velocity inversion}. In such a scenario, the group delay incurred by the ST wave packet is zero when traversing a bilayer formed of equal-thickness layers of indices $n_{1}$ and $n_{2}$ when the group index in the first layer is $\widetilde{n}_{1}\!=\!n_{1}-n_{2}$. We denote this scenario \textit{group-delay cancellation} [Fig.~\ref{Fig:Consequences}(d)]. Considering incidence from an ambient medium of index $n_{0}$, then a wave packet having group index $\widetilde{n}_{0}\!=\!1-n_{1}n_{2}$ will incur a delay $\tau_{1}\!=\!(n_{1}-n_{2})L/c$ when normally incident on a layer of thickness $L$ and index $n_{1}$, and a delay $\tau_{2}\!=\!(n_{2}-n_{1})L/c\!=\!-\tau_{1}$ when incident on a layer of equal thickness and index $n_{2}$.

A similar effect can be realized for unequal layer thicknesses $L_{1}$ and $L_{2}$ of refractive indices $n_{1}$ and $n_{2}$, respectively. Group-delay cancellation occurs when $L_{1}\widetilde{n}_{1}+L_{2}\widetilde{n}_{2}\!=\!0$, which necessitates the use of an incident wave packet whose group index is $\widetilde{n}_{1}\!=\!\tfrac{n_{1}^{2}-n_{2}^{2}}{L_{2}n_{1}+L_{1}n_{2}}L_{2}$, for which the transmitted wave packet has a group index $\widetilde{n}_{2}\!=\!\tfrac{n_{2}^{2}-n_{1}^{2}}{L_{2}n_{1}+L_{1}n_{2}}L_{1}\!=\!-\tfrac{L_{1}}{L_{2}}\widetilde{n}_{1}$.  When incident from free space, the required group index is $\widetilde{n}_{0}\!=\!1-n_{1}n_{2}\tfrac{L_{1}n_{1}-L-{2}n_{2}}{L_{1}n_{2}+L_{2}n_{1}}$.

\subsection{GVD for large bandwidths}

The parabolic approximation in Eq.~\ref{Eq:BasebandParabola}, leading to the law of refraction in Eq.~\ref{Eq:BasebandLawOfRefraction}, was derived under the assumption of narrow spatial and temporal bandwidths. Violating this assumption (e.g., due to a broad spatio-temporal bandwidth) does \textit{not} invalidate Eq.~\ref{Eq:BasebandLawOfRefraction}, which still dictates the change in $\widetilde{v}$. However, in addition to the change in $\widetilde{v}$, the transmitted ST wave packet experiences group velocity dispersion (GVD) in the second medium. To estimate this GVD, we obtain the axial wave number $k_{z_{2}}$ in the second medium in terms of $n_{1}$, $n_{2}$, and $\widetilde{n}_{1}$. Subtracting the equations for the light-cones in the two media $k_{x}^{2}+k_{z_{1}}^{2}\!=\!n_{1}^{2}(\tfrac{\omega}{c})^{2}$ and $k_{x}^{2}+k_{z_{2}}^{2}\!=\!n_{2}^{2}(\tfrac{\omega}{c})^{2}$ and then substituting for $k_{z_{1}}\!=\!n_{1}k_{\mathrm{o}}+\tfrac{\Omega}{c}\widetilde{n}_{1}$, we obtain this exact relationship:
\begin{equation}
k_{z_{2}}^{2}=n_{2}^{2}k_{\mathrm{o}}^{2}+2(n_{2}^{2}-n_{1}^{2}+n_{1}\widetilde{n}_{1})k_{\mathrm{o}}\tfrac{\Omega}{c}+(n_{2}^{2}-n_{1}^{2}+\widetilde{n}_{1}^{2})(\tfrac{\Omega}{c})^{2}.
\end{equation}
We expand $k_{z_{2}}$ in the vicinity of $k_{x}\!=\!0$ ($\omega\!=\!\omega_{\mathrm{o}}$):
\begin{equation}
k_{z_{2}}(\Omega)\approx k_{z_{2}}(\Omega\!=\!0)+\frac{dk_{z_{2}}}{d\Omega}\big|_{\Omega\!=\!0}\Omega+\frac{1}{2}\frac{d^{2}k_{z_{2}}}{d\Omega^{2}}\big|_{\Omega\!=\!0}\Omega^{2}+\ldots
\end{equation}
The zeroth-order term is $k_{z_{2}}(0)\!=\!n_{2}k_{\mathrm{o}}$, whereas the first-order term is $\tfrac{dk_{z_{2}}}{d\Omega}|_{\Omega\!=\!0}\!=\!\tfrac{n_{2}^{2}-n_{1}^{2}+n_{1}\widetilde{n}_{1}}{n_{2}c}\!=\!\tfrac{\widetilde{n}_{2}}{c}$, corresponding to the law of refraction in Eq.~\ref{Eq:BasebandLawOfRefraction}. Defining the GVD parameter $k_{2}\!=\!\tfrac{d^{2}k_{z_{2}}}{d\Omega^{2}}|_{\Omega\!=\!0}$, we have:
\begin{equation}
k_{2}\!=\!\frac{(n_{2}^{2}\!-\!n_{1}^{2})-(\widetilde{n}_{2}^{2}\!-\!\widetilde{n}_{1}^{2})}{n_{2}k_{\mathrm{o}}c^{2}}\!=\!\frac{1}{k_{\mathrm{o}}c^{2}}\,\frac{1}{n_{2}}(n_{1}\!-\!\widetilde{n}_{1})^{2}\left(1\!-\frac{n_{1}^{2}}{n_{2}^{2}}\right).
\end{equation}
Therefore, GVD \textit{increases} with: (1) the index contrast between $n_{1}$ and $n_{2}$; (2) the deviation of the group index in the first medium $\widetilde{n}_{1}$ from the luminal condition; and (3) the wavelength $\lambda_{\mathrm{o}}$. The GVD is normal ($k_{2}\!>\!0$) when $n_{2}\!>\!n_{1}$ and is anomalous ($k_{2}\!<\!0$) when $n_{2}\!<\!n_{1}$. The value of $k_{2}$ (for fixed $n_{1}$ and $\widetilde{n}_{1}$) reaches a maximum value when $n_{2}\!=\!\sqrt{3}n_{1}$ given by
\begin{equation}
k_{2}^{(\mathrm{max})}\!=\!\frac{1}{k_{\mathrm{o}}c^{2}}\frac{2}{3\sqrt{3}}\frac{1}{n_{1}}(n_{1}-\widetilde{n}_{1})^{2}.
\end{equation}

\section{Normal incidence of X-waves}

\subsection{Law of refraction}

Starting from Eq.~\ref{Eq:ExactXWave} for a X-wave, we can easily identify the refractive invariant quantity $n^{2}-\widetilde{n}^{2}\!=\!(\tfrac{k_{x}}{\omega/c})^{2}$. This invariant for X-waves is \textit{exact} and independent of bandwidth because the spectral projections are straight lines. We can thus formulate the following law of refraction at normal incidence for X-waves: 
\begin{equation}\label{Eq:XWaveLawOfRefraction}
n_{1}^{2}-\widetilde{n}_{1}^{2}=n_{2}^{2}-\widetilde{n}_{2}^{2},
\end{equation}
which is plotted in Fig.~\ref{Fig:XWavesAndSidebandPlot} (dashed curve). Because only positive-valued superluminal group velocities are accessible $\widetilde{n}\!<\!n$, the axes for $\theta_{1}$ and $\theta_{2}$ in Fig.~\ref{Fig:XWavesAndSidebandPlot} extend only to $90^{\circ}$. 

Practically, we expect that $\widetilde{n}_{1}\!\approx\!n_{1}$ and $\widetilde{n}_{2}\!\approx\!n_{2}$; i.e., there is always only a minute deviation in group velocity from the luminal limit. In fact, previous experiments have reported group velocities of X-waves in free space of $\widetilde{v}\!\approx\!1.0002c$ \cite{Bowlan09OL,Bonaretti09OE,Kuntz09PRA} (confirmed also in \cite{Yessenov19PRA}). This is a fundamental limitation for X-waves that cannot be overcome within paraxial optics. Indeed, if we take $k_{x}\!=\!nk_{\mathrm{o}}\delta$ (where $\delta\!\ll\!1$), then the deviation from the luminal condition is $\Delta\widetilde{v}\!=\!\widetilde{v}-\tfrac{c}{n}\!\approx\!\tfrac{\delta^{2}}{2n}c$. For $\delta\!=\!0.1$, which approaches the limit of paraxial optics, we have $\Delta\widetilde{v}\!\approx\!0.005c$ in free space. Realizing a X-wave whose group velocity deviates appreciably from $c$ therefore requires operating deep within the non-paraxial regime.

\subsection{Consequences of the law of refraction}

It is important to note that normal incidence of a X-wave is only \textit{nominal}. In contrast to baseband ST wave packets where the plane-wave component at $\omega\!=\!\omega_{\mathrm{o}}$ is normally incident on the interface, the component at $\omega_{\mathrm{o}}$ for a X-wave is \textit{not} associated with $k_{x}\!=\!0$, and is \textit{not} normally incident; see Fig.~\ref{Fig:XandSidebandIncidence}(a). Indeed, the angle of incidence $\phi_{1}'$ of $\omega_{\mathrm{o}}$ in a medium of index $n_{1}$ is given by $\cos{\phi_{1}}\!=\!\widetilde{n}_{1}/n_{1}$ [Fig.~\ref{Fig:XandSidebandIncidence}(a)]. For this effective incident angle to reach the critical condition for total internal reflection (when $n_{1}\!>\!n_{2}$) requires an extreme non-paraxial scenario, and we can safely ignore this possibility.  Finally, as a result of the exact formulation of the law of refraction for X-waves (independently of bandwidth) GVD never arises for X-waves upon refraction at normal incidence. Although the law of refraction here is a quadratic equation, the relationship between $\theta_{1}$ and $\theta_{2}$ is monotonic over the allowable range, with $\widetilde{n}_{2}$ reaching a minimum value of $\widetilde{n}_{2}\!=\!\sqrt{n_{2}^{2}-n_{1}^{2}}$ at $\widetilde{n}_{1}\!=\!0$ ($\theta_{1}\!=\!90^{\circ}$) [Fig.~\ref{Fig:XWavesAndSidebandPlot}]. 

\subsubsection{The luminal wave packet remains luminal}

Because the luminal limit for a X-wave is a plane-wave pulse ($k_{x}\!=\!0$), the invariance of $k_{x}$ implies that a luminal X-wave remains luminal. Indeed, the luminal limits for baseband ST wave packets and X-waves coincide.

\subsubsection{Normal refraction}

\begin{figure}[t!]
  \begin{center}
  \includegraphics[width=8.6cm]{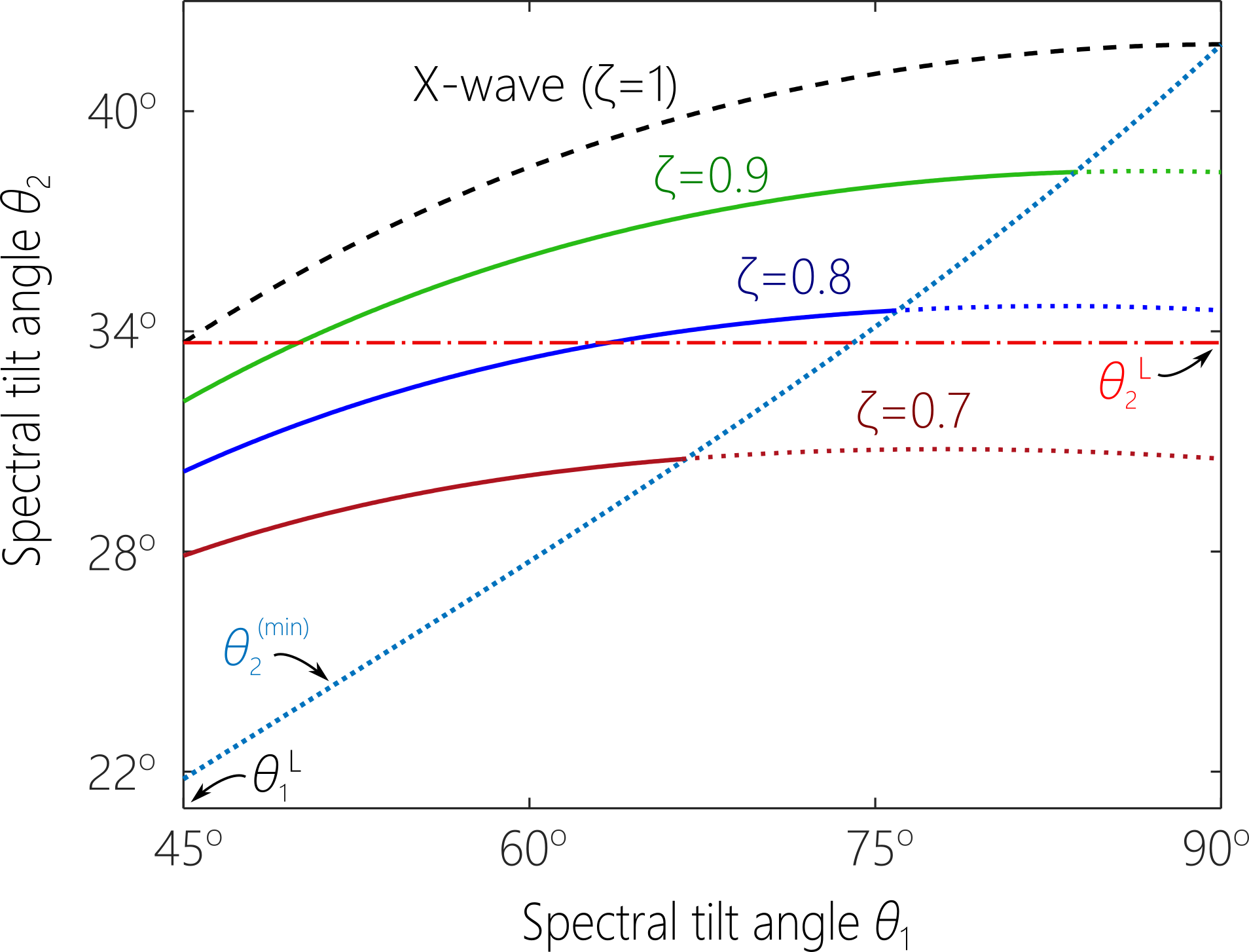}
  \end{center}
  \caption{Law of refraction for X-waves (dashed curve; Eq.~\ref{Eq:XWaveLawOfRefraction}) and sideband ST wave packets (solid curves; Eq.\ref{Eq:SidebandLawOfRefraction}) for different values of $\zeta$, all at normal incidence. We use $n_{1}\!=\!1$ and $n_{2}\!=\!1.5$, but the main features of this plot are generic for all values of $n_{1}$ and $n_{2}$. The axes for $\theta_{1}$ and $\theta_{2}$ extend to only $90^{\circ}$ -- and not to $180^{\circ}$ as in Fig.~\ref{Fig:BasebandPlot}(c) -- because negative superluminal velocities are forbidden. Furthermore, the subluminal regime is also excluded. The dotted curve corresponds to Eq.~\ref{Eq:MimimumTheta2} and dictates the minimal value for $\theta_{2}$ achievable for a given value of $\theta_{1}$ that can be reached by varying the value of $\zeta$. At $\theta_{1}\!=\!90^{\circ}$, this curve meets up with the X-wave limit, and at the luminal limit here $\theta_{1}\!=\!45^{\circ}$, we have $\theta_{2}^{(\mathrm{min})}\!\approx\!21.8^{\circ}$.}
  \label{Fig:XWavesAndSidebandPlot}
\end{figure}

The law of refraction for normally incident X-waves in Eq.~\ref{Eq:XWaveLawOfRefraction} indicates that their refraction is always \textit{normal}: the group velocity always drops when traveling to a higher-index medium and increases when traveling to a lower-index medium, just as with conventional pulses. In other words, if $n_{2}\!>\!n_{1}$, then $\widetilde{n}_{2}\!>\widetilde{n}_{1}$. This is seen in Fig.~\ref{Fig:XWavesAndSidebandPlot} by noting that the entire curve representing the law of refraction lies \textit{below} the diagonal $\theta_{1}\!=\!\theta_{2}$. As a result, the phenomena described above for baseband ST wave packets, including group-velocity invariance and anomalous refraction, do \textit{not} occur with X-waves. Moreover, group-velocity inversion does not occur because X-waves are restricted to positive-valued group velocities.

\begin{figure*}[t!]
  \begin{center}
  \includegraphics[width=16.5cm]{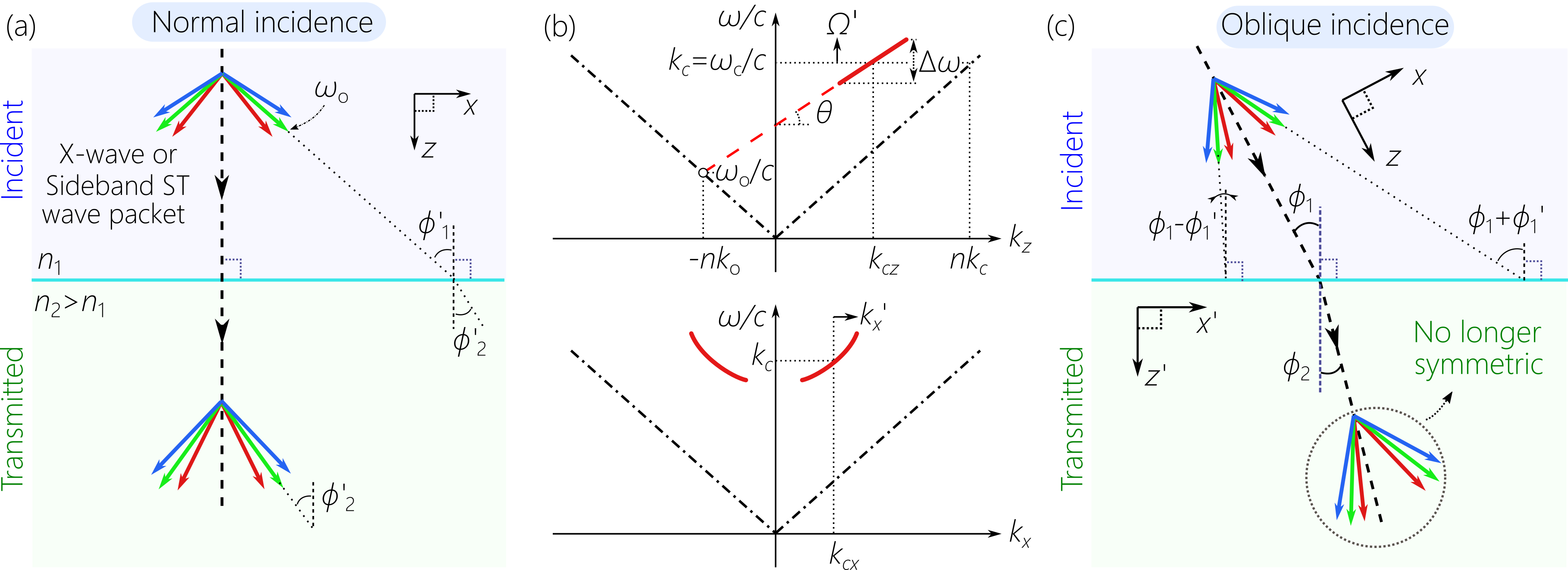}
  \end{center}
  \caption{(a) Schematic of a normally incident X-wave or sideband ST wave packet at a planar interface between two dielectrics of refractive indices $n_{1}$ and $n_{2}$. The angle $\phi_{1}'$ is that made by the carrier frequency $\omega_{\mathrm{o}}$ for a X-wave at the interface when the wave packet is nominally at normal incidence. (b) Spectral projections onto the $(k_{z},\tfrac{\omega}{c})$ and $(k_{x},\tfrac{\omega}{c})$ planes for a sideband ST wave packet with narrow bandwidth. We identify here the parameters used in the text to formulate the law of refraction in Eq.~\ref{Eq:SidebandLawOfRefraction}. (c) Same as (a) for oblique incidence at an angle $\phi_{1}$ with respect to the normal to the interface.}
  \label{Fig:XandSidebandIncidence}
\end{figure*}

\section{Normal incidence of sideband space-time wave packets}

\subsection{Law of refraction}

It is more difficult to formulate a law of refraction for sideband ST wave packets (e.g., FWMs) because the spatial frequencies in the vicinity of $k_{x}\!=\!0$ are forbidden on physical grounds, thereby precluding the parabolic approximation used for baseband ST wave packets. Instead, the spectral plane $\mathcal{P}_{\mathrm{S}}(\theta)$ intersects with the light-cone in a conic section \textit{and} the support domain corresponds to large values of $k_{x}$, which can even be comparable to $k_{z}$ (i.e., non-paraxial); see Fig.~\ref{Fig:ThreeClasses}(c).

Nevertheless, an approximation can be formulated in the limit of a \textit{narrow} spectral bandwidth. Unlike baseband ST wave packets and X-waves, the frequency $\omega_{\mathrm{o}}$ does \textit{not} belong to the spectrum of sideband ST wave packets. Assume a center frequency $\omega_{\mathrm{c}}\!>\!\omega_{\mathrm{o}}$ and narrow bandwidth $\Delta\omega\!\ll\!\omega_{\mathrm{c}}$. Define the wave number $k_{\mathrm{c}}\!=\!\tfrac{\omega_{\mathrm{c}}}{c}$, and the transverse and axial components $k_{\mathrm{c}x}$ and $k_{\mathrm{c}z}$, respectively, with $k_{\mathrm{c}x}^{2}+k_{\mathrm{c}z}^{2}\!=\!n^{2}k_{\mathrm{c}}^{2}$; see Fig.~\ref{Fig:XandSidebandIncidence}(b). The narrow temporal bandwidth implies also a narrow spatial bandwidth $\Delta k_{x}\!\ll\!k_{\mathrm{c}x}$. Starting from the exact conic section in Eq.~\ref{Eq:ExactSideband}, expanding the temporal frequency $\omega\!=\!\omega_{\mathrm{c}}+\Omega'$ and the spatial frequency $k_{x}\!=\!k_{\mathrm{c}x}+k_{x}'$, with $\Omega'\!\ll\!\omega_{\mathrm{c}}$ and $k_{x}'\!\ll\!k_{\mathrm{c}x}$, and retaining terms that are first-order in $\Omega'$ and $k_{x}'$ yields a refractive invariant,
\begin{equation}
(n+\widetilde{n})(n-\zeta\widetilde{n})=\frac{k_{x}'}{\Omega'/c}\,\frac{k_{\mathrm{c}x}}{k_{\mathrm{c}}},
\end{equation}
where $\zeta\!=\!1-\tfrac{\omega_{\mathrm{o}}}{\omega_{\mathrm{c}}}\!<\!1$ is a dimensionless parameter close to unity. This refractive invariant is the slope of the spectral projection of the ST wave packet onto the $(k_{x},\tfrac{\omega}{c})$-plane. From this invariant quantity, we obtain a law of refraction for sideband ST wave packets at normal incidence [Fig.~\ref{Fig:XWavesAndSidebandPlot}]:
\begin{equation}\label{Eq:SidebandLawOfRefraction}
(n_{1}+\widetilde{n}_{1})(n_{1}-\zeta\widetilde{n}_{1})=(n_{2}+\widetilde{n}_{2})(n_{2}-\zeta\widetilde{n}_{2}).
\end{equation}

The parameter $\zeta$ is limited to the range $\tfrac{n}{n+\widetilde{n}}\!<\!\zeta\!<\!1$. At its maximum value $\zeta\!\rightarrow\!1$ ($\omega_{\mathrm{c}}\!\gg\!\omega_{\mathrm{o}}$ or $k_{\mathrm{o}}\!\rightarrow\!0$), the wave packet reverts to a X-wave, and the law of refraction in Eq.~\ref{Eq:SidebandLawOfRefraction} simplifies to that in Eq.~\ref{Eq:XWaveLawOfRefraction} for X-waves. The minimal value $\zeta\!=\!\tfrac{n}{n+\widetilde{n}}$ corresponds to $k_{\mathrm{c}z}\!=\!0$ and $k_{\mathrm{c}x}\!=\!nk_{\mathrm{c}}$, whereupon $\omega\!=\!(1+\tfrac{n}{\widetilde{n}})\omega_{\mathrm{o}}$. In general, the axial wave number is a fraction of the total wave number in the medium $k_{\mathrm{c}z}\!=\!\eta nk_{\mathrm{c}}$, with $0\!<\!\eta\!<\!1$. It is easy to show that 
\begin{equation}
\zeta=\frac{1+\eta}{1+\frac{\widetilde{n}}{n}},
\end{equation}
which implies an additional constraint $\eta\!<\!\tfrac{\widetilde{n}}{n}$ in the superluminal regime (it is always satisfied in the subluminal regime). Therefore, remaining within paraxial limits ($\eta\!\rightarrow\!1$ and $\zeta\!\rightarrow\!1$) necessitates staying close to the luminal condition ($\widetilde{n}\!\rightarrow\!1$).

To the best of our knowledge, there have been no reported measurements of the group velocity of sideband ST wave packets. Indeed, the conditions for producing a deviation from $c$ here are even more stringent than those for X-waves. If the center frequency of the spectrum is $\omega_{\mathrm{c}}$, then the deviation in group velocity from the luminal condition in free space is $\tfrac{\Delta\widetilde{v}}{c}\!=\!\tfrac{\widetilde{v}-c}{c}\!=\!\tfrac{\Delta\widetilde{v}_{\mathrm{X}}}{c}-2(1-\zeta)\tfrac{1}{\eta}$, where $\eta\!=\!k_{z}/k_{\mathrm{c}}$ and $\Delta\widetilde{v}_{\mathrm{X}}$ is the deviation for a X-wave of similar spatial bandwidth. As such, a sideband ST wave packet will be even closer to the luminal limit than a X-wave of similar spatial bandwidth.

\subsection{Consequences of the law of refraction}

Although the equation representing the law of refraction for sideband ST wave packets is quadratic, nevertheless the relationship between $\theta_{1}$ and $\theta_{2}$ is monotonic because the constraint $\zeta\!>\!\tfrac{n_{1}}{n_{1}+\widetilde{n}_{1}}$ restricts the valid domain in $(\theta_{1},\theta_{2})$-space to that above the dotted curve ($\zeta\!=\!\tfrac{n}{n+\widetilde{n}}$) in Fig.~\ref{Fig:XWavesAndSidebandPlot}.

Similarly to X-waves, the central frequency $\omega_{\mathrm{c}}$ for a sideband ST wave packet is \textit{not} normally incident on the interface  [Fig.~\ref{Fig:XandSidebandIncidence}(a)], but instead makes an angle $\phi_{1}'$ with respect to the normal to the interface, where
\begin{equation}\label{Eq:PhiDashForSideband}
\cos{\phi_{1}'}=\frac{\widetilde{n}_{1}}{n_{1}}-\left(1+\frac{\widetilde{n}_{1}}{n_{1}}\right)\frac{\omega_{\mathrm{o}}}{\omega_{\mathrm{c}}};
\end{equation}
the second term in this expression vanishes for X-waves. The minimal value of $\zeta$ (maximum value for $\tfrac{\omega_{\mathrm{o}}}{\omega_{\mathrm{c}}}$) leads to $\cos{\phi_{1}'}\!=\!0$ ($\phi_{1}'\!=\!90^{\circ}$), which is far from the paraxial regime as expected. We emphasize that the findings discussed below for sideband ST wave packets are only valid for narrow bandwidths. 

\subsubsection{Group-velocity invariance}

The law of refraction in Eq.~\ref{Eq:SidebandLawOfRefraction} has two solutions for group-velocity invariance $\widetilde{n}_{2}\!=\!\widetilde{n}_{1}\!=\!\widetilde{n}_{\mathrm{th}}$. The first is the trivial solution $n_{1}\!=\!n_{2}$; the second is at a negative-valued group index $\widetilde{n}_{\mathrm{th}}\!=\!-\frac{n_{1}+n_{2}}{1-\zeta}\!<\!0$, which is excluded. Therefore, similarly to X-waves, sideband ST wave packets do not exhibit group-velocity invariance and their refraction is therefore always \textit{normal} [Fig.~\ref{Fig:XWavesAndSidebandPlot}]. Furthermore, because only positive-valued group velocities can be realized, group-velocity inversion and group-delay cancellation are also precluded

\subsubsection{Bridging the subluminal-superluminal barrier}

Unlike baseband ST wave packets and X-waves, where a luminal incident wave packet remains luminal, a luminal sideband ST wave packet (i.e., a FWM) does \textit{not} remain luminal upon transmission. This is confirmed by setting $\widetilde{n}_{1}\!=\!n_{1}$  (a luminal FWM) in Eq.~\ref{Eq:SidebandLawOfRefraction}, whereupon $\tfrac{\widetilde{n}_{2}}{n_{2}}\!=\!\tfrac{1+\Delta}{2\zeta}-\tfrac{1-\Delta}{2}$; where $\Delta^{2}\!=\!1-8\frac{\zeta(1-\zeta)}{(1+\zeta)^{2}}\tfrac{n_{1}^{2}}{n_{2}^{2}}$. As such $\widetilde{n}_{2}\!\neq\!n_{2}$; i.e., the transmitted wave packet is \textit{not} luminal. Indeed, assuming $n_{2}\!>\!n_{1}$, we have $\widetilde{n}_{2}\!>\!n_{2}$ so that the transmitted wave packet is \textit{subluminal}, as expected in the normal refraction regime. As $\widetilde{v}_{1}$ increases (thus becoming superluminal), $\widetilde{v}_{2}$ also increases until it reaches the luminal value of $\widetilde{v}_{2}\!=\!c/n_{2}$ when $\widetilde{n}_{1}\!=\!\tfrac{1}{2\zeta}\{n_{1}(1-\zeta)+\sqrt{n_{1}^{2}(1+\zeta)^{2}-8n_{2}^{2}\zeta(1-\zeta)}\}$. 

As $\widetilde{v}_{1}$ increases further, $\widetilde{v}_{2}$ also increases until it formally reaches a maximum value when $\widetilde{n}_{1}\!=\!n_{1}\tfrac{1-\zeta}{2\zeta}$, followed by a drop in $\widetilde{v}_{2}$ with further increase in $\widetilde{v}_{1}$ beyond this value. However, this regime is inaccessible because of the abovementioned limit set on the minimum value of $\zeta$. Indeed, reaching the peak in $\widetilde{v}_{2}$ as $\widetilde{v}_{1}$ increases requires $\zeta\!=\!\tfrac{n_{1}}{n_{1}+2\widetilde{n}_{1}}\!<\!\zeta_{\mathrm{min}}\!=\!\tfrac{n_{1}}{n_{1}+\widetilde{n}_{1}}$, which is not admissible [Fig.~\ref{Fig:XWavesAndSidebandPlot}]. The minimal value of $\zeta$ imposes a minimum achievable value $\theta_{2}^{(\mathrm{min})}$ at any $\theta_{1}$ given by
\begin{equation}\label{Eq:MimimumTheta2}
\widetilde{n}_{2}^{(\mathrm{max})}\!=\!\cot{\theta_{2}^{(\mathrm{min})}}\!=\!\frac{n_{2}}{2}\frac{\widetilde{n}_{1}}{n_{1}}\!+\!\sqrt{n_{2}^{2}\left(1\!+\!\frac{1}{2}\frac{\widetilde{n}_{1}}{n_{1}}\right)^{2}\!-\!n_{1}^{2}\left(1\!+\!\frac{\widetilde{n}_{1}}{n_{1}}\right)}.
\end{equation}
The dotted curve in Fig.~\ref{Fig:XWavesAndSidebandPlot} corresponds to this equation. It can be easily confirmed that setting $\widetilde{n}_{1}\!=\!0$ ($\theta_{1}\!=\!90^{\circ}$) results in $\widetilde{n}_{2}^{(\mathrm{max})}\!=\!\sqrt{n_{2}^{2}-n_{1}^{2}}$, which coincides with the X-waves limit; and setting the luminal limit for FWMs $\widetilde{n}_{1}\!=\!n_{1}$ results in $\widetilde{n}_{2}^{(\mathrm{max})}\!=\!0.5n_{2}+\sqrt{2.25n_{2}^{2}-2n_{1}^{2}}$.

\section{Oblique incidence of baseband space-time wave packets}

Although the transverse momentum is invariant across the planar interface, at oblique incidence the transverse component of the wave vector $k_{x}$ with respect to the propagation axis of the ST wave packet is \textit{not} parallel to the interface [Fig.~\ref{Fig:NormalAndOblique}(b)]. If $k_{x_{1}}\!=\!n_{1}k\sin{\alpha_{1}}$, where $\alpha_{1}$ is measured with respect to the incident propagation direction, $k\!=\!\tfrac{\omega}{c}$, and $\phi_{1}$ is the angle of incidence with respect to the normal to the interface, then the invariant transverse momentum with respect to the interface is $k_{x_{1}}'\!=\!n_{1}k\sin{(\phi_{1}+\alpha_{1})}$, whereupon $k_{x_{1}}'\!=\!k_{x_{2}}'$; see Fig.~\ref{Fig:NormalAndOblique}(b). We place no restriction on the value of $\phi_{1}$, but we retain the small-bandwidth approximation for the ST wave packets, $\alpha\!\ll\!1$ ($\sin{\alpha}\!\approx\!\alpha$ and $\cos{\alpha}\!\approx\!1$), so that $k_{x}'\!\approx\!k_{x}\cos{\phi}+nk\sin{\phi}$. Because $n_{1}\sin{\phi_{1}}\!=\!n_{2}\sin{\phi_{2}}$ (Snell's law) and $k_{x_{1}}'\!=\!k_{x_{2}}'$, we have $k_{x1}\cos{\phi_{1}}\!\approx\!k_{x2}\cos{\phi_{2}}$. Referring to Eq.~\ref{Eq:BasebandParabola}, the invariance of $\omega$ and $k_{x}\cos{\phi}$ across the interface imply that the quantity $n(n-\widetilde{n})\cos^{2}{\phi}$ is the new refractive invariant at oblique incidence, from which we formulate the modified law:
\begin{equation}\label{Eq:Oblique_LawForBaseband}
n_{1}(n_{1}-\widetilde{n}_{1})\cos^{2}{\phi_{1}}=n_{2}(n_{2}-\widetilde{n}_{2})\cos^{2}{\phi_{2}}.
\end{equation}
This relationship reverts to Eq.~\ref{Eq:BasebandLawOfRefraction} at normal incidence.

\subsection{Consequences of the law of refraction}

\begin{figure}[t!]
  \begin{center}
  \includegraphics[width=8.2cm]{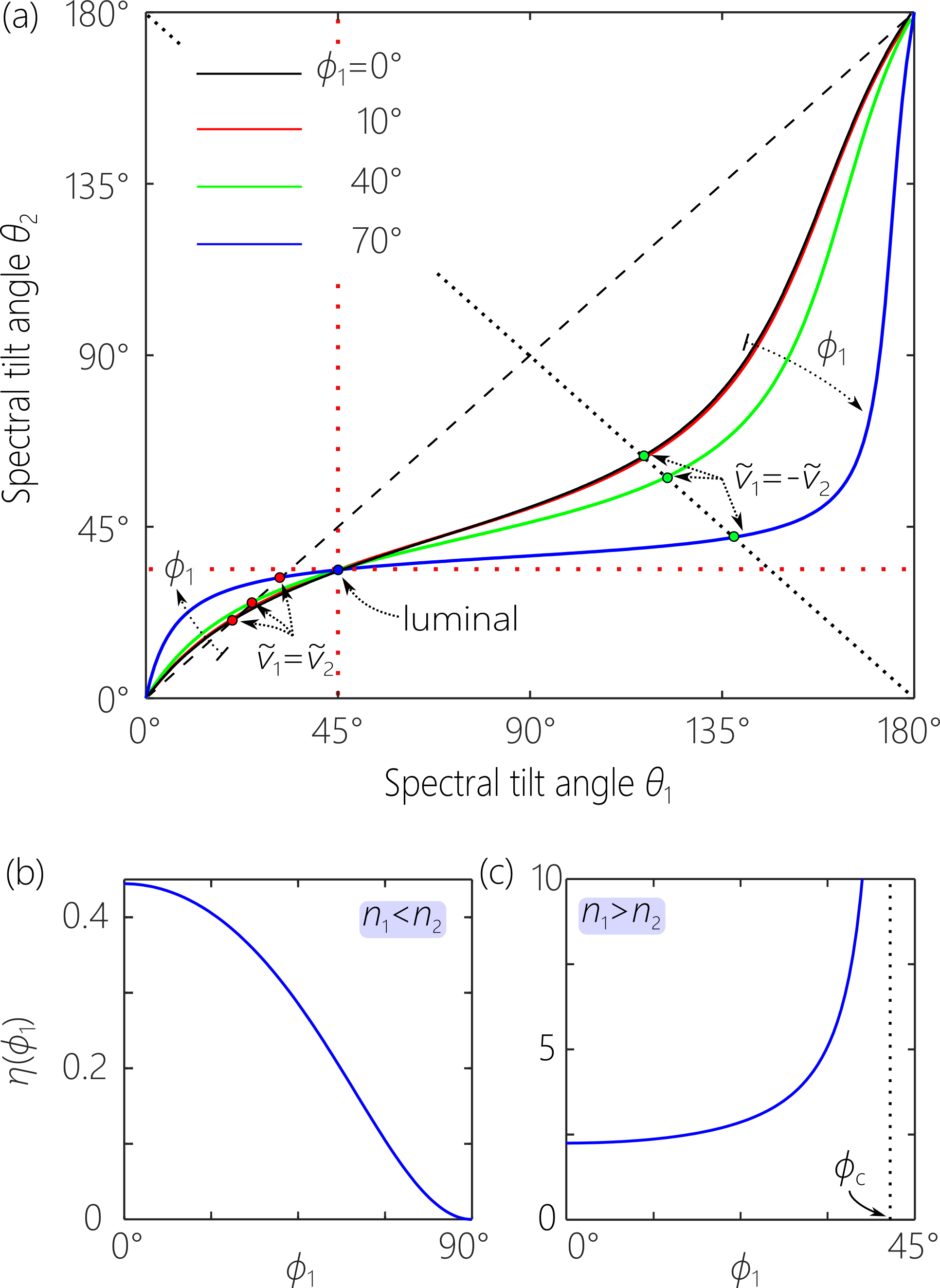}
  \end{center}
  \caption{(a) Law of refraction for baseband ST wave packets at oblique incidence (Eq.~\ref{Eq:Oblique_LawForBaseband}) for $n_{1}\!=\!1$ and $n_{2}\!=\!1.5$, but the features of these plots are generic for all values of $n_{1}$ and $n_{2}$. We highlight the shift in the conditions for group-velocity invariance $\widetilde{v}_{2}\!=\!\widetilde{v}_{1}$ and group-velocity inversion $\widetilde{v}_{2}\!=\!-\widetilde{v}_{1}$ with increasing $\phi_{1}$. (b) Plot of $\eta(\phi_{1})$ in Eq.~\ref{Eq:gFunction} for $n_{1}\!=\!1$ and $n_{2}\!=\!1.5$ and (c) for $n_{2}\!=\!1$ and $n_{1}\!=\!1.5$; $\phi_{\mathrm{c}}$ is the critical angle $\sin{\phi_{\mathrm{c}}}\!=\!\tfrac{n_{2}}{n_{1}}$.}
  \label{Fig:BasebandObliquePlot}
\end{figure}

The consequences of normal-incidence refraction for baseband ST wave packets are retained at oblique incidence after the appropriate modifications. First, luminal wave packets remain luminal at oblique incidence; i.e., $\widetilde{n}_{1}\!=\!n_{1}$ still entails that $\widetilde{n}_{2}\!=\!n_{2}$. Second, the group index for group-velocity invariance $\widetilde{n}_{1}\!=\!\widetilde{n}_{2}\!=\!\widetilde{n}_{\mathrm{th}}$ is
\begin{equation}
\widetilde{n}_{\mathrm{th}}(\phi_{1})=\frac{n_{1}+n_{2}}{1+\frac{n_{1}}{n_{2}}\sin^{2}{\phi_{1}}}<\widetilde{n}_{\mathrm{th}}(0)=n_{1}+n_{2}.
\end{equation}
In other words, the invariant group velocity is higher at oblique incidence. The onset of anomalous refraction can thus be tuned by changing the angle of incidence [Fig.~\ref{Fig:BasebandObliquePlot}(a)]. Third, the condition for group-velocity inversion $\widetilde{n}_{1}\!=\!-\widetilde{n}_{2}$ at oblique incidence is:
\begin{equation}\label{Eq:ObliqueGroupVelocityInversion}
\widetilde{n}_{1}(\phi_{1})=\frac{n_{1}-n_{2}}{1-\frac{n_{1}}{n_{2}}\sin^{2}{\phi_{1}}}.
\end{equation}
When $n_{1}\!<\!n_{2}$, we have $\widetilde{n}_{1}(\phi_{1})\!<\!\widetilde{n}_{1}(0)\!=\!n_{1}-n_{2}$, so that the incident group velocity at which group-velocity inversion occurs increases. The opposite occurs when $n_{1}\!>\!n_{2}$, whereupon $\widetilde{n}_{1}(\phi_{1})\!>\!\widetilde{n}_{1}(0)$ and the value of the incident group velocity at which group-velocity inversion occurs decreases. The condition in Eq.~\ref{Eq:ObliqueGroupVelocityInversion} guarantees group-velocity inversion but \textit{not} group-delay cancellation because at oblique incidence the propagation distances are not equal in two equal-thickness layers, which necessitates modifying Eq.~\ref{Eq:ObliqueGroupVelocityInversion} to accommodate the path-length difference in the two layers.

\subsection{Dependence of the transmitted-wave-packet group velocity on angle of incidence} 

A unique consequence of refraction at oblique incidence is that the group velocity in the second medium can be tuned by changing the angle of incidence $\phi_{1}$ (at fixed $\widetilde{n}_{1}$). By rewriting the law of refraction in Eq.~\ref{Eq:Oblique_LawForBaseband}, $\tfrac{\widetilde{n}_{2}}{n_{2}}=1-\left(1-\tfrac{\widetilde{n}_{1}}{n_{1}}\right)\eta(\phi_{1})$, $\phi_{1}$ dictates $\widetilde{n}_{2}$ through the function $g(\phi_{1})$ given by:
\begin{equation}\label{Eq:gFunction}
\eta(\phi_{1})=\frac{\cos^{2}{\phi_{1}}}{\frac{n_{2}^{2}}{n_{1}^{2}}-\sin^{2}{\phi_{1}}}.
\end{equation}
The behavior of this function is quite distinct in the two cases of $n_{1}\!<\!n_{2}$ and $n_{1}\!>\!n_{2}$. When $n_{1}\!<\!n_{2}$, $\eta(\phi_{1})$ \textit{decreases} monotonically from a maximum value of $n_{1}^{2}/n_{2}^{2}$ at $\phi_{1}\!=\!0$ to a minimum value of $\eta(\phi_{1})\!=\!0$ at $\phi_{1}\!=\!90^{\circ}$ [Fig.~\ref{Fig:BasebandObliquePlot}(b)]. On the other hand, when $n_{1}\!>\!n_{2}$, $\eta(\phi_{1})$ \textit{increases} monotonically from a minimum value of $n_{1}^{2}/n_{2}^{2}$ at $\phi_{1}\!=\!0$, and then becomes unbounded when approaching the critical angle $\sin{\phi_{\mathrm{c}}}\!=\!n_{2}/n_{1}$ [Fig.~\ref{Fig:BasebandObliquePlot}(c)].

We can now describe the behavior of $\widetilde{v}_{2}$ when tuning the angle of incidence $\phi_{1}$. First, when $n_{2}\!>\!n_{1}$ and the wave packet is \textit{subluminal}, $\widetilde{v}_{2}$ \textit{increases} with $\phi_{1}$ in a higher-index medium. On the other hand, for \textit{superluminal} wave packets, $\widetilde{v}_{2}$ \textit{decreases} with $\phi_{1}$ in a higher-index medium. These trends are reversed when $n_{2}\!<\!n_{1}$. For \textit{subluminal} wave packets, $\widetilde{v}_{2}$ \textit{decreases} with $\phi_{1}$ in a lower-index medium -- approaching zero at the critical angle $\phi_{\mathrm{c}}$. On the other hand, for \textit{superluminal} wave packets, $\widetilde{v}_{2}$ \textit{increases} with $\phi_{1}$ in a lower-index medium, formally reaching infinity at $\phi_{\mathrm{c}}$. This unique behavior leads to an intriguing scenario for blind optical synchronization suggested in Ref.~\cite{Bhaduri20NP}, and studied in more detail and realized in paper (III). Additionally, this behavior underlies the recent demonstration of `isochronous' ST wave packets that traverse a planar slab at a fixed group delay at any angle of incidence \cite{Motz21arxiv}.

\section{Oblique incidence of X-waves and sideband ST wave packets}

The analysis provided above for baseband ST wave packets at oblique incidence does \textit{not} extend to X-waves or sideband ST wave packets. The reason is that their spatial spectra are \textit{not} centered at $k_{x}\!=\!0$. Instead, the spatial spectrum of the incident wave packet is centered on a non-zero incident angle $\pm\phi_{1}'$, where $\cos{\phi_{1}'}\!=\!\tfrac{\widetilde{n}_{1}}{n_{1}}$ for X-waves and is given by Eq.~\ref{Eq:PhiDashForSideband} for sideband ST wave packets. At normal incidence [Fig.~\ref{Fig:XandSidebandIncidence}(a)], this does not effect the symmetry of the transmitted wave packet. However, at oblique incidence [Fig.~\ref{Fig:XandSidebandIncidence}(c)], the two bands of the spatial spectrum (comprising the positive and negative spatial frequencies) impinge on the interface at \textit{different} angles $\phi_{1}\pm\phi_{1}'$, and thus also refract at different angles. Consequently, the two bands of the spatial spectrum in the second medium \textit{are no longer symmetric with respect to the direction of propagation}. This can also be appreciated by noting that the component of the wave vector transverse to the interface (and thus invariant across it) is $k_{\mathrm{T}}\!=\!k_{x_{1}}\cos{\phi_{1}}+k_{z_{1}}\sin{\phi_{1}}\!=\!k_{x_{2}}\cos{\phi_{2}}+k_{z_{2}}\sin{\phi_{2}}$. Consequently, the transverse and angular wave numbers ($k_{x}$ and $k_{z}$, respectively) get mixed, thereby fundamentally changing the spatio-temporal spectral structure. This fundamental feature of the refraction of sideband ST wave packets was first recognized in Ref.~\cite{Donnelly97IEEE} for the special case of FWMs. The ST wave packet is no longer propagation invariant. Only when the wave packets are extremely close to the luminal limit $\widetilde{n}_{1}\!\rightarrow\!n_{1}$ can this effect be potentially ignored; however, in this case no significant deviation from conventional refraction is observed. As such, we cannot obtain useful laws of refraction for X-waves nor for sideband ST wave packets at oblique incidence. 

\section{Discussion and conclusion}

In Table~\ref{Table:normal} we summarize the main refraction characteristics for baseband, X-wave, and sideband ST wave packets at normal incidence, including the refractive invariant that allows formulating a law of refraction, the condition for group-velocity invariance, the existence of an anomalous refraction regime, the condition for group-velocity inversion, and whether the subluminal-superluminal barrier can be bridged. Table~\ref{Table:ObliqueVsNormal} provides the modifications to the refractive phenomena associated with baseband ST wave packets at oblique incidence with respect to their normal-incidence counterparts. Finally, Table~\ref{Table:LargerOrSmallerIndex} summarizes the refraction of obliquely incident baseband ST wave packets when $n_{1}\!<\!n_{2}$ or $n_{1}\!>\!n_{2}$.

We have studied ST wave packets transmitted across a planar interface between two dielectrics.  Our formulation is based solely on an examination of the spectral support domain on the light-cone. The reflected ST wave packet retains the group velocity of the incident wave packet, with potential changes in the spatio-temporal profile due to the polarization-dependent Fresnel coefficients. Despite potential changes in the complex field amplitudes at each wavelength, and thus changes in the profile (an external DoF), its unique propagation characteristics are nevertheless dictated solely by the underlying spatio-temporal spectral structure (an internal DoF). In the analysis presented here, we have not dealt with the impact of the Fresnel coefficients on the refraction of ST wave packets. The complex field amplitudes along the invariant spectral projection onto the $(k_{x},\tfrac{\omega}{c})$-plane will change upon transmission according to the Fresnel coefficients. At normal incidence, the changes will be symmetric around $k_{x}\!=\!0$ and independent of wavelength for non-dispersive materials; at oblique incidence, the changes are \textit{not} symmetric around $k_{x}\!=\!0$ and are wavelength-dependent even in non-dispersive materials. It is crucial to emphasize that the impact of the Fresnel coefficients is to change the complex field amplitudes and thus the spatio-temporal profile of the wave packet. This change is polarization-dependent, and is more significant at oblique incidence and for large spatial bandwidths. The Fresnel coefficients have no impact on the spectral support domain or the spectral tilt angle. Therefore, the laws of refraction we have formulated here are \textit{independent} of polarization. 

\begin{table}[t!]
\caption{Summary of the refraction characteristics for baseband, X-wave, and sideband ST wave packets at normal incidence.}\label{Table:normal}
\begin{tabular}{|p{1.8cm}||p{1.7cm}|p{1.3cm}|p{2.0cm}|}
 \hline
  & baseband & X-wave & sideband \\
 \hline\hline
 Invariant & $n(n-\widetilde{n})$ & $n^{2}-\widetilde{n}^{2}$ & $(n\!+\!\widetilde{n})(n\!-\!\zeta\widetilde{n})$ \\
 \hline
$\widetilde{v}$-invariance& $\widetilde{n}_{\mathrm{th}}\!=\!n_{1}\!+\!n_{2}$ & No & No\\
 \hline
  Anomalous & $\widetilde{n}_{1}\!>\!\widetilde{n}_{\mathrm{th}}$ & No & No \\
 \hline
 $\widetilde{v}$-inversion& $\widetilde{n}_{1}\!=\!n_{1}\!-\!n_{2}$ & No & No\\
 \hline
 Bridging gap & No & No & Yes \\
 \hline
\end{tabular}\end{table}

\begin{table}[t!]
\caption{Summary of the refraction characteristics for obliquely incident baseband ST wave packets; $\phi_{1}$ is the angle of incidence.}\label{Table:ObliqueVsNormal}
\begin{tabular}{|p{2cm}||p{2.3cm}|p{2.9cm}|}\hline
  & Normal & Oblique \\ \hline\hline
 Invariant & $n(n-\widetilde{n})$ & $n(n-\widetilde{n})\cos^{2}{\phi}$ \\ \hline 
$\widetilde{v}$-invariance& $\widetilde{n}_{\mathrm{th}}(0)\!=\!n_{1}\!+\!n_{2}$ & $\widetilde{n}_{\mathrm{th}}(\phi_{1})\!=\!\frac{n_{1}+n_{2}}{1+\frac{n_{1}}{n_{2}}\sin^{2}{\phi_{1}}}$\\ \hline
  Anomalous & $\widetilde{n}_{1}\!>\!\widetilde{n}_{\mathrm{th}}(0)$ & $\widetilde{n}_{1}\!>\!\widetilde{n}_{\mathrm{th}}(\phi)$ \\  \hline
 $\widetilde{v}$-inversion& $\widetilde{n}_{1}(0)\!=\!n_{1}\!-\!n_{2}$ & $\widetilde{n}_{1}(\phi_{1})\!=\!\frac{n_{1}-n_{2}}{1-\frac{n_{1}}{n_{2}}\sin^{2}{\phi_{1}}}$\\ \hline
 Bridging gap & No & No \\ \hline
\end{tabular}\end{table}

\begin{table}[t!]
\caption{Summary of the differences in refraction characteristics for obliquely incident baseband ST wave packets when $n_{1}\!<\!n_{2}$ or $n_{1}\!>\!n_{2}$; $\phi_{1}$ is the angle of incidence.}\label{Table:LargerOrSmallerIndex}
\begin{tabular}{|p{1.9cm}||p{2.8cm}|p{2.8cm}|}\hline
  & $n_{1}\!<\!n_{2}$ & $n_{1}\!>\!n_{2}$ \\ \hline\hline
 $\widetilde{v}$-invariance & no change & no change \\ \hline
 $\widetilde{v}$-inversion & $\widetilde{n}_{1}(\phi_{1})<\widetilde{n}_{1}(0)$ & $\widetilde{n}_{1}(\phi_{1})>\widetilde{n}_{1}(0)$ \\ \hline
sublum.-$\widetilde{v}_{1}$ & $\widetilde{v}_{2}$ increases with $\phi_{1}$ & $\widetilde{v}_{2}$ decreases with $\phi_{1}$ \\ \hline
superlum.-$\widetilde{v}_{1}$ & $\widetilde{v}_{2}$ decreases with $\phi_{1}$ & $\widetilde{v}_{2}$ increases with $\phi_{1}$ \\ \hline
\end{tabular}\end{table}

In conclusion, we have established laws of refraction for the three classes of ST wave packets (baseband, X-waves, and sideband) at normal incidence. These laws govern the change in an \textit{internal} degree of freedom (the spectral tilt angle) for the transmitted wave packet across a planar interface between two non-dispersive, homogeneous, isotropic dielectrics, and are formulated by identifying a new refractive optical invariant: the spectral curvature. The group velocity of the transmitted wave packet depends not only on the refractive index of the two media, but also on the group velocity of the incident wave packet. Consequently, baseband ST wave packets exhibit fascinating refractive phenomena, such as anomalous refraction, and group-velocity invariance and inversion. At oblique incidence, a law of refraction is formulated for baseband ST wave packets, which shows that the group velocity of the transmitted wave packet depends also on the angle of incidence. The phenomena associated with normal incidence of baseband ST wave packets are retained at oblique incidence after appropriate modifications. Laws of oblique refraction are \textit{not} formulated for X-waves and sideband ST wave packets, whose structures are fundamentally changed after refraction at oblique incidence. Understanding the refraction of ST wave packets at planar interfaces is crucial for studying their interaction with photonic devices, such as waveguides \cite{Zamboni01PRE,Zamboni03PRE,Shiri20NC,Kibler21PRL,Bejot21arxiv} and Fabry-P{\'e}rot cavities \cite{Shabahang17SR,Villinger19unpubl,Shiri20OL,Shiri20APLP}.

In paper (II) and paper (III) of this sequence we provide experimental confirmation of these theoretical predictions regarding baseband ST wave packets.

\section*{Funding}
U.S. Office of Naval Research (ONR) contract N00014-17-1-2458.

\vspace{2mm}
\noindent
\textbf{Disclosures.} The authors declare no conflicts of interest.

\bibliography{diffraction}

\begin{thebibliography}{84}%
\makeatletter
\providecommand \@ifxundefined [1]{%
 \@ifx{#1\undefined}
}%
\providecommand \@ifnum [1]{%
 \ifnum #1\expandafter \@firstoftwo
 \else \expandafter \@secondoftwo
 \fi
}%
\providecommand \@ifx [1]{%
 \ifx #1\expandafter \@firstoftwo
 \else \expandafter \@secondoftwo
 \fi
}%
\providecommand \natexlab [1]{#1}%
\providecommand \enquote  [1]{``#1''}%
\providecommand \bibnamefont  [1]{#1}%
\providecommand \bibfnamefont [1]{#1}%
\providecommand \citenamefont [1]{#1}%
\providecommand \href@noop [0]{\@secondoftwo}%
\providecommand \href [0]{\begingroup \@sanitize@url \@href}%
\providecommand \@href[1]{\@@startlink{#1}\@@href}%
\providecommand \@@href[1]{\endgroup#1\@@endlink}%
\providecommand \@sanitize@url [0]{\catcode `\\12\catcode `\$12\catcode
  `\&12\catcode `\#12\catcode `\^12\catcode `\_12\catcode `\%12\relax}%
\providecommand \@@startlink[1]{}%
\providecommand \@@endlink[0]{}%
\providecommand \url  [0]{\begingroup\@sanitize@url \@url }%
\providecommand \@url [1]{\endgroup\@href {#1}{\urlprefix }}%
\providecommand \urlprefix  [0]{URL }%
\providecommand \Eprint [0]{\href }%
\providecommand \doibase [0]{http://dx.doi.org/}%
\providecommand \selectlanguage [0]{\@gobble}%
\providecommand \bibinfo  [0]{\@secondoftwo}%
\providecommand \bibfield  [0]{\@secondoftwo}%
\providecommand \translation [1]{[#1]}%
\providecommand \BibitemOpen [0]{}%
\providecommand \bibitemStop [0]{}%
\providecommand \bibitemNoStop [0]{.\EOS\space}%
\providecommand \EOS [0]{\spacefactor3000\relax}%
\providecommand \BibitemShut  [1]{\csname bibitem#1\endcsname}%
\let\auto@bib@innerbib\@empty
\bibitem [{\citenamefont {Donnelly}\ and\ \citenamefont
  {Ziolkowski}(1993)}]{Donnelly93ProcRSLA}%
  \BibitemOpen
  \bibfield  {author} {\bibinfo {author} {\bibfnamefont {R.}~\bibnamefont
  {Donnelly}}\ and\ \bibinfo {author} {\bibfnamefont {R.~W.}\ \bibnamefont
  {Ziolkowski}},\ }\href@noop {} {\bibfield  {journal} {\bibinfo  {journal}
  {Proc. R. Soc. Lond. A}\ }\textbf {\bibinfo {volume} {440}},\ \bibinfo
  {pages} {541} (\bibinfo {year} {1993})}\BibitemShut {NoStop}%
\bibitem [{\citenamefont {Longhi}(2004)}]{Longhi04OE}%
  \BibitemOpen
  \bibfield  {author} {\bibinfo {author} {\bibfnamefont {S.}~\bibnamefont
  {Longhi}},\ }\href@noop {} {\bibfield  {journal} {\bibinfo  {journal} {Opt.
  Express}\ }\textbf {\bibinfo {volume} {12}},\ \bibinfo {pages} {935}
  (\bibinfo {year} {2004})}\BibitemShut {NoStop}%
\bibitem [{\citenamefont {Saari}\ and\ \citenamefont
  {Reivelt}(2004)}]{Saari04PRE}%
  \BibitemOpen
  \bibfield  {author} {\bibinfo {author} {\bibfnamefont {P.}~\bibnamefont
  {Saari}}\ and\ \bibinfo {author} {\bibfnamefont {K.}~\bibnamefont
  {Reivelt}},\ }\href@noop {} {\bibfield  {journal} {\bibinfo  {journal} {Phys.
  Rev. E}\ }\textbf {\bibinfo {volume} {69}},\ \bibinfo {pages} {036612}
  (\bibinfo {year} {2004})}\BibitemShut {NoStop}%
\bibitem [{\citenamefont {Kondakci}\ and\ \citenamefont
  {Abouraddy}(2016)}]{Kondakci16OE}%
  \BibitemOpen
  \bibfield  {author} {\bibinfo {author} {\bibfnamefont {H.~E.}\ \bibnamefont
  {Kondakci}}\ and\ \bibinfo {author} {\bibfnamefont {A.~F.}\ \bibnamefont
  {Abouraddy}},\ }\href@noop {} {\bibfield  {journal} {\bibinfo  {journal}
  {Opt. Express}\ }\textbf {\bibinfo {volume} {24}},\ \bibinfo {pages} {28659}
  (\bibinfo {year} {2016})}\BibitemShut {NoStop}%
\bibitem [{\citenamefont {Parker}\ and\ \citenamefont
  {Alonso}(2016)}]{Parker16OE}%
  \BibitemOpen
  \bibfield  {author} {\bibinfo {author} {\bibfnamefont {K.~J.}\ \bibnamefont
  {Parker}}\ and\ \bibinfo {author} {\bibfnamefont {M.~A.}\ \bibnamefont
  {Alonso}},\ }\href@noop {} {\bibfield  {journal} {\bibinfo  {journal} {Opt.
  Express}\ }\textbf {\bibinfo {volume} {24}},\ \bibinfo {pages} {28669}
  (\bibinfo {year} {2016})}\BibitemShut {NoStop}%
\bibitem [{\citenamefont {Kondakci}\ and\ \citenamefont
  {Abouraddy}(2017)}]{Kondakci17NP}%
  \BibitemOpen
  \bibfield  {author} {\bibinfo {author} {\bibfnamefont {H.~E.}\ \bibnamefont
  {Kondakci}}\ and\ \bibinfo {author} {\bibfnamefont {A.~F.}\ \bibnamefont
  {Abouraddy}},\ }\href@noop {} {\bibfield  {journal} {\bibinfo  {journal}
  {Nat. Photon.}\ }\textbf {\bibinfo {volume} {11}},\ \bibinfo {pages} {733}
  (\bibinfo {year} {2017})}\BibitemShut {NoStop}%
\bibitem [{\citenamefont {Kondakci}\ \emph {et~al.}(2019)\citenamefont
  {Kondakci}, \citenamefont {Alonso},\ and\ \citenamefont
  {Abouraddy}}]{Kondakci19OL}%
  \BibitemOpen
  \bibfield  {author} {\bibinfo {author} {\bibfnamefont {H.~E.}\ \bibnamefont
  {Kondakci}}, \bibinfo {author} {\bibfnamefont {M.~A.}\ \bibnamefont
  {Alonso}}, \ and\ \bibinfo {author} {\bibfnamefont {A.~F.}\ \bibnamefont
  {Abouraddy}},\ }\href@noop {} {\bibfield  {journal} {\bibinfo  {journal}
  {Opt. Lett.}\ }\textbf {\bibinfo {volume} {44}},\ \bibinfo {pages} {2645}
  (\bibinfo {year} {2019})}\BibitemShut {NoStop}%
\bibitem [{\citenamefont {Hall}\ \emph {et~al.}(2021)\citenamefont {Hall},
  \citenamefont {Yessenov},\ and\ \citenamefont {Abouraddy}}]{Hall21OL}%
  \BibitemOpen
  \bibfield  {author} {\bibinfo {author} {\bibfnamefont {L.~A.}\ \bibnamefont
  {Hall}}, \bibinfo {author} {\bibfnamefont {M.}~\bibnamefont {Yessenov}}, \
  and\ \bibinfo {author} {\bibfnamefont {A.~F.}\ \bibnamefont {Abouraddy}},\
  }\href@noop {} {\bibfield  {journal} {\bibinfo  {journal} {Opt. Lett.}\
  }\textbf {\bibinfo {volume} {46}},\ \bibinfo {pages} {1672} (\bibinfo {year}
  {2021})}\BibitemShut {NoStop}%
\bibitem [{\citenamefont {Reivelt}\ and\ \citenamefont
  {Saari}(2003)}]{Reivelt03arxiv}%
  \BibitemOpen
  \bibfield  {author} {\bibinfo {author} {\bibfnamefont {K.}~\bibnamefont
  {Reivelt}}\ and\ \bibinfo {author} {\bibfnamefont {P.}~\bibnamefont
  {Saari}},\ }\href@noop {} {\bibfield  {journal} {\bibinfo  {journal}
  {arxiv:physics/0309079}\ } (\bibinfo {year} {2003})}\BibitemShut {NoStop}%
\bibitem [{\citenamefont {Kiselev}(2007)}]{Kiselev07OS}%
  \BibitemOpen
  \bibfield  {author} {\bibinfo {author} {\bibfnamefont {A.~P.}\ \bibnamefont
  {Kiselev}},\ }\href@noop {} {\bibfield  {journal} {\bibinfo  {journal} {Opt.
  Spectrosc.}\ }\textbf {\bibinfo {volume} {102}},\ \bibinfo {pages} {603}
  (\bibinfo {year} {2007})}\BibitemShut {NoStop}%
\bibitem [{\citenamefont {Hern\'andez-Figueroa}\ \emph
  {et~al.}(2008)\citenamefont {Hern\'andez-Figueroa}, \citenamefont {Recami},\
  and\ \citenamefont {Zamboni-Rached}}]{FigueroaBook8}%
  \BibitemOpen
  \bibinfo {editor} {\bibfnamefont {H.~E.}\ \bibnamefont
  {Hern\'andez-Figueroa}}, \bibinfo {editor} {\bibfnamefont {E.}~\bibnamefont
  {Recami}}, \ and\ \bibinfo {editor} {\bibfnamefont {M.}~\bibnamefont
  {Zamboni-Rached}},\ eds.,\ \href@noop {} {\emph {\bibinfo {title} {Localized
  waves}}}\ (\bibinfo  {publisher} {Wiley-Interscience},\ \bibinfo {year}
  {2008})\BibitemShut {NoStop}%
\bibitem [{\citenamefont {Turunen}\ and\ \citenamefont
  {Friberg}(2010)}]{Turunen10PO}%
  \BibitemOpen
  \bibfield  {author} {\bibinfo {author} {\bibfnamefont {J.}~\bibnamefont
  {Turunen}}\ and\ \bibinfo {author} {\bibfnamefont {A.~T.}\ \bibnamefont
  {Friberg}},\ }\href@noop {} {\bibfield  {journal} {\bibinfo  {journal} {Prog.
  Opt.}\ }\textbf {\bibinfo {volume} {54}},\ \bibinfo {pages} {1} (\bibinfo
  {year} {2010})}\BibitemShut {NoStop}%
\bibitem [{\citenamefont {Hern\'andez-Figueroa}\ \emph
  {et~al.}(2014)\citenamefont {Hern\'andez-Figueroa}, \citenamefont {Recami},\
  and\ \citenamefont {Zamboni-Rached}}]{FigueroaBook14}%
  \BibitemOpen
  \bibinfo {editor} {\bibfnamefont {H.~E.}\ \bibnamefont
  {Hern\'andez-Figueroa}}, \bibinfo {editor} {\bibfnamefont {E.}~\bibnamefont
  {Recami}}, \ and\ \bibinfo {editor} {\bibfnamefont {M.}~\bibnamefont
  {Zamboni-Rached}},\ eds.,\ \href@noop {} {\emph {\bibinfo {title}
  {Non-diffracting Waves}}}\ (\bibinfo  {publisher} {Wiley-VCH},\ \bibinfo
  {year} {2014})\BibitemShut {NoStop}%
\bibitem [{\citenamefont {Salo}\ and\ \citenamefont
  {Salomaa}(2001)}]{Salo01JOA}%
  \BibitemOpen
  \bibfield  {author} {\bibinfo {author} {\bibfnamefont {J.}~\bibnamefont
  {Salo}}\ and\ \bibinfo {author} {\bibfnamefont {M.~M.}\ \bibnamefont
  {Salomaa}},\ }\href@noop {} {\bibfield  {journal} {\bibinfo  {journal} {J.
  Opt. A}\ }\textbf {\bibinfo {volume} {3}},\ \bibinfo {pages} {366} (\bibinfo
  {year} {2001})}\BibitemShut {NoStop}%
\bibitem [{\citenamefont {Brittingham}(1983)}]{Brittingham83JAP}%
  \BibitemOpen
  \bibfield  {author} {\bibinfo {author} {\bibfnamefont {J.~N.}\ \bibnamefont
  {Brittingham}},\ }\href@noop {} {\bibfield  {journal} {\bibinfo  {journal}
  {J. Appl. Phys.}\ }\textbf {\bibinfo {volume} {54}},\ \bibinfo {pages} {1179}
  (\bibinfo {year} {1983})}\BibitemShut {NoStop}%
\bibitem [{\citenamefont {Lu}\ and\ \citenamefont
  {Greenleaf}(1992)}]{Lu92IEEEa}%
  \BibitemOpen
  \bibfield  {author} {\bibinfo {author} {\bibfnamefont {J.-Y.}\ \bibnamefont
  {Lu}}\ and\ \bibinfo {author} {\bibfnamefont {J.~F.}\ \bibnamefont
  {Greenleaf}},\ }\href@noop {} {\bibfield  {journal} {\bibinfo  {journal}
  {IEEE Trans. Ultrason. Ferroelec. Freq. Control}\ }\textbf {\bibinfo {volume}
  {39}},\ \bibinfo {pages} {19} (\bibinfo {year} {1992})}\BibitemShut {NoStop}%
\bibitem [{\citenamefont {Saari}\ and\ \citenamefont
  {Reivelt}(1997)}]{Saari97PRL}%
  \BibitemOpen
  \bibfield  {author} {\bibinfo {author} {\bibfnamefont {P.}~\bibnamefont
  {Saari}}\ and\ \bibinfo {author} {\bibfnamefont {K.}~\bibnamefont
  {Reivelt}},\ }\href@noop {} {\bibfield  {journal} {\bibinfo  {journal} {Phys.
  Rev. Lett.}\ }\textbf {\bibinfo {volume} {79}},\ \bibinfo {pages} {4135}
  (\bibinfo {year} {1997})}\BibitemShut {NoStop}%
\bibitem [{\citenamefont {Zamboni-Rached}\ \emph {et~al.}(2002)\citenamefont
  {Zamboni-Rached}, \citenamefont {Recami},\ and\ \citenamefont
  {Hern{\'a}ndez-Figueroa}}]{Zamboni02EPJD}%
  \BibitemOpen
  \bibfield  {author} {\bibinfo {author} {\bibfnamefont {M.}~\bibnamefont
  {Zamboni-Rached}}, \bibinfo {author} {\bibfnamefont {E.}~\bibnamefont
  {Recami}}, \ and\ \bibinfo {author} {\bibfnamefont {H.~E.}\ \bibnamefont
  {Hern{\'a}ndez-Figueroa}},\ }\href@noop {} {\bibfield  {journal} {\bibinfo
  {journal} {Eur. Phys. J. D}\ }\textbf {\bibinfo {volume} {21}},\ \bibinfo
  {pages} {217} (\bibinfo {year} {2002})}\BibitemShut {NoStop}%
\bibitem [{\citenamefont {Recami}\ \emph {et~al.}(2003)\citenamefont {Recami},
  \citenamefont {Zamboni-Rached}, \citenamefont {N{\'o}brega},\ and\
  \citenamefont {Dartora}}]{Recami03IEEEJSTQE}%
  \BibitemOpen
  \bibfield  {author} {\bibinfo {author} {\bibfnamefont {E.}~\bibnamefont
  {Recami}}, \bibinfo {author} {\bibfnamefont {M.}~\bibnamefont
  {Zamboni-Rached}}, \bibinfo {author} {\bibfnamefont {K.~Z.}\ \bibnamefont
  {N{\'o}brega}}, \ and\ \bibinfo {author} {\bibfnamefont {C.~A.}\ \bibnamefont
  {Dartora}},\ }\href@noop {} {\bibfield  {journal} {\bibinfo  {journal} {IEEE
  J. Sel. Top. Quantum Electron.}\ }\textbf {\bibinfo {volume} {9}},\ \bibinfo
  {pages} {59} (\bibinfo {year} {2003})}\BibitemShut {NoStop}%
\bibitem [{\citenamefont {Valtna}\ \emph {et~al.}(2007)\citenamefont {Valtna},
  \citenamefont {Reivelt},\ and\ \citenamefont {Saari}}]{Valtna07OC}%
  \BibitemOpen
  \bibfield  {author} {\bibinfo {author} {\bibfnamefont {H.}~\bibnamefont
  {Valtna}}, \bibinfo {author} {\bibfnamefont {K.}~\bibnamefont {Reivelt}}, \
  and\ \bibinfo {author} {\bibfnamefont {P.}~\bibnamefont {Saari}},\
  }\href@noop {} {\bibfield  {journal} {\bibinfo  {journal} {Opt. Commun.}\
  }\textbf {\bibinfo {volume} {278}},\ \bibinfo {pages} {1} (\bibinfo {year}
  {2007})}\BibitemShut {NoStop}%
\bibitem [{\citenamefont {Zamboni-Rached}(2009)}]{Zamboni09PRA}%
  \BibitemOpen
  \bibfield  {author} {\bibinfo {author} {\bibfnamefont {M.}~\bibnamefont
  {Zamboni-Rached}},\ }\href@noop {} {\bibfield  {journal} {\bibinfo  {journal}
  {Phys. Rev. A}\ }\textbf {\bibinfo {volume} {79}},\ \bibinfo {pages} {013816}
  (\bibinfo {year} {2009})}\BibitemShut {NoStop}%
\bibitem [{\citenamefont {Zamboni-Rached}\ and\ \citenamefont
  {Recami}(2008)}]{Zamboni08PRA}%
  \BibitemOpen
  \bibfield  {author} {\bibinfo {author} {\bibfnamefont {M.}~\bibnamefont
  {Zamboni-Rached}}\ and\ \bibinfo {author} {\bibfnamefont {E.}~\bibnamefont
  {Recami}},\ }\href@noop {} {\bibfield  {journal} {\bibinfo  {journal} {Phys.
  Rev. A}\ }\textbf {\bibinfo {volume} {77}},\ \bibinfo {pages} {033824}
  (\bibinfo {year} {2008})}\BibitemShut {NoStop}%
\bibitem [{\citenamefont {Zapata-Rodr{\'i}guez}\ and\ \citenamefont
  {Porras}(2006)}]{Zapata06OL}%
  \BibitemOpen
  \bibfield  {author} {\bibinfo {author} {\bibfnamefont {C.~J.}\ \bibnamefont
  {Zapata-Rodr{\'i}guez}}\ and\ \bibinfo {author} {\bibfnamefont {M.~A.}\
  \bibnamefont {Porras}},\ }\href@noop {} {\bibfield  {journal} {\bibinfo
  {journal} {Opt. Lett.}\ }\textbf {\bibinfo {volume} {31}},\ \bibinfo {pages}
  {3532} (\bibinfo {year} {2006})}\BibitemShut {NoStop}%
\bibitem [{\citenamefont {Porras}(2017)}]{Porras17OL}%
  \BibitemOpen
  \bibfield  {author} {\bibinfo {author} {\bibfnamefont {M.~A.}\ \bibnamefont
  {Porras}},\ }\href@noop {} {\bibfield  {journal} {\bibinfo  {journal} {Opt.
  Lett.}\ }\textbf {\bibinfo {volume} {42}},\ \bibinfo {pages} {4679} (\bibinfo
  {year} {2017})}\BibitemShut {NoStop}%
\bibitem [{\citenamefont {Efremidis}(2017)}]{Efremidis17OL}%
  \BibitemOpen
  \bibfield  {author} {\bibinfo {author} {\bibfnamefont {N.~K.}\ \bibnamefont
  {Efremidis}},\ }\href@noop {} {\bibfield  {journal} {\bibinfo  {journal}
  {Opt. Lett.}\ }\textbf {\bibinfo {volume} {42}},\ \bibinfo {pages} {5038}
  (\bibinfo {year} {2017})}\BibitemShut {NoStop}%
\bibitem [{\citenamefont {Wong}\ and\ \citenamefont
  {Kaminer}(2017{\natexlab{a}})}]{Wong17ACSP1}%
  \BibitemOpen
  \bibfield  {author} {\bibinfo {author} {\bibfnamefont {L.~J.}\ \bibnamefont
  {Wong}}\ and\ \bibinfo {author} {\bibfnamefont {I.}~\bibnamefont {Kaminer}},\
  }\href@noop {} {\bibfield  {journal} {\bibinfo  {journal} {ACS Photon.}\
  }\textbf {\bibinfo {volume} {4}},\ \bibinfo {pages} {1131} (\bibinfo {year}
  {2017}{\natexlab{a}})}\BibitemShut {NoStop}%
\bibitem [{\citenamefont {Wong}\ and\ \citenamefont
  {Kaminer}(2017{\natexlab{b}})}]{Wong17ACSP2}%
  \BibitemOpen
  \bibfield  {author} {\bibinfo {author} {\bibfnamefont {L.~J.}\ \bibnamefont
  {Wong}}\ and\ \bibinfo {author} {\bibfnamefont {I.}~\bibnamefont {Kaminer}},\
  }\href@noop {} {\bibfield  {journal} {\bibinfo  {journal} {ACS Photon.}\
  }\textbf {\bibinfo {volume} {4}},\ \bibinfo {pages} {2257} (\bibinfo {year}
  {2017}{\natexlab{b}})}\BibitemShut {NoStop}%
\bibitem [{\citenamefont {Sainte-Marie}\ \emph {et~al.}(2017)\citenamefont
  {Sainte-Marie}, \citenamefont {Gobert},\ and\ \citenamefont
  {Qu{\'e}r{\'e}}}]{SaintMarie17Optica}%
  \BibitemOpen
  \bibfield  {author} {\bibinfo {author} {\bibfnamefont {A.}~\bibnamefont
  {Sainte-Marie}}, \bibinfo {author} {\bibfnamefont {O.}~\bibnamefont
  {Gobert}}, \ and\ \bibinfo {author} {\bibfnamefont {F.}~\bibnamefont
  {Qu{\'e}r{\'e}}},\ }\href@noop {} {\bibfield  {journal} {\bibinfo  {journal}
  {Optica}\ }\textbf {\bibinfo {volume} {4}},\ \bibinfo {pages} {1298}
  (\bibinfo {year} {2017})}\BibitemShut {NoStop}%
\bibitem [{\citenamefont {Porras}(2018)}]{PorrasPRA18}%
  \BibitemOpen
  \bibfield  {author} {\bibinfo {author} {\bibfnamefont {M.~A.}\ \bibnamefont
  {Porras}},\ }\href@noop {} {\bibfield  {journal} {\bibinfo  {journal} {Phys.
  Rev. A}\ }\textbf {\bibinfo {volume} {97}},\ \bibinfo {pages} {063803}
  (\bibinfo {year} {2018})}\BibitemShut {NoStop}%
\bibitem [{\citenamefont {Wong}\ \emph {et~al.}(2020)\citenamefont {Wong},
  \citenamefont {Christodoulides},\ and\ \citenamefont {Kaminer}}]{Wong20AS}%
  \BibitemOpen
  \bibfield  {author} {\bibinfo {author} {\bibfnamefont {L.~J.}\ \bibnamefont
  {Wong}}, \bibinfo {author} {\bibfnamefont {D.~N.}\ \bibnamefont
  {Christodoulides}}, \ and\ \bibinfo {author} {\bibfnamefont {I.}~\bibnamefont
  {Kaminer}},\ }\href@noop {} {\bibfield  {journal} {\bibinfo  {journal} {Adv.
  Sci.}\ }\textbf {\bibinfo {volume} {7}},\ \bibinfo {pages} {1903377}
  (\bibinfo {year} {2020})}\BibitemShut {NoStop}%
\bibitem [{\citenamefont {Kibler}\ and\ \citenamefont
  {B{\'e}jot}(2021)}]{Kibler21PRL}%
  \BibitemOpen
  \bibfield  {author} {\bibinfo {author} {\bibfnamefont {B.}~\bibnamefont
  {Kibler}}\ and\ \bibinfo {author} {\bibfnamefont {P.}~\bibnamefont
  {B{\'e}jot}},\ }\href@noop {} {\bibfield  {journal} {\bibinfo  {journal}
  {Phys. Rev. Lett.}\ }\textbf {\bibinfo {volume} {126}},\ \bibinfo {pages}
  {023902} (\bibinfo {year} {2021})}\BibitemShut {NoStop}%
\bibitem [{\citenamefont {Shen}\ \emph {et~al.}(2021)\citenamefont {Shen},
  \citenamefont {Zdagkas}, \citenamefont {Papasimakis},\ and\ \citenamefont
  {Zheludev}}]{Shen21PRR}%
  \BibitemOpen
  \bibfield  {author} {\bibinfo {author} {\bibfnamefont {Y.}~\bibnamefont
  {Shen}}, \bibinfo {author} {\bibfnamefont {A.}~\bibnamefont {Zdagkas}},
  \bibinfo {author} {\bibfnamefont {N.}~\bibnamefont {Papasimakis}}, \ and\
  \bibinfo {author} {\bibfnamefont {N.~I.}\ \bibnamefont {Zheludev}},\
  }\href@noop {} {\bibfield  {journal} {\bibinfo  {journal} {Phys. Rev. Res.}\
  }\textbf {\bibinfo {volume} {3}},\ \bibinfo {pages} {013236} (\bibinfo {year}
  {2021})}\BibitemShut {NoStop}%
\bibitem [{\citenamefont {B{\'e}jot}\ and\ \citenamefont
  {Kibler}(2021)}]{Bejot21arxiv}%
  \BibitemOpen
  \bibfield  {author} {\bibinfo {author} {\bibfnamefont {P.}~\bibnamefont
  {B{\'e}jot}}\ and\ \bibinfo {author} {\bibfnamefont {B.}~\bibnamefont
  {Kibler}},\ }\href@noop {} {\bibfield  {journal} {\bibinfo  {journal}
  {arXiv:2103.11620}\ } (\bibinfo {year} {2021})}\BibitemShut {NoStop}%
\bibitem [{\citenamefont {Froula}\ \emph {et~al.}(2018)\citenamefont {Froula},
  \citenamefont {Turnbull}, \citenamefont {Davies}, \citenamefont {Kessler},
  \citenamefont {Haberberger}, \citenamefont {Palastro}, \citenamefont {Bahk},
  \citenamefont {Begishev}, \citenamefont {Boni}, \citenamefont {Bucht},
  \citenamefont {Katz},\ and\ \citenamefont {Shaw}}]{Froula18NP}%
  \BibitemOpen
  \bibfield  {author} {\bibinfo {author} {\bibfnamefont {D.~H.}\ \bibnamefont
  {Froula}}, \bibinfo {author} {\bibfnamefont {D.}~\bibnamefont {Turnbull}},
  \bibinfo {author} {\bibfnamefont {A.~S.}\ \bibnamefont {Davies}}, \bibinfo
  {author} {\bibfnamefont {T.~J.}\ \bibnamefont {Kessler}}, \bibinfo {author}
  {\bibfnamefont {D.}~\bibnamefont {Haberberger}}, \bibinfo {author}
  {\bibfnamefont {J.~P.}\ \bibnamefont {Palastro}}, \bibinfo {author}
  {\bibfnamefont {S.-W.}\ \bibnamefont {Bahk}}, \bibinfo {author}
  {\bibfnamefont {I.~A.}\ \bibnamefont {Begishev}}, \bibinfo {author}
  {\bibfnamefont {R.}~\bibnamefont {Boni}}, \bibinfo {author} {\bibfnamefont
  {S.}~\bibnamefont {Bucht}}, \bibinfo {author} {\bibfnamefont
  {J.}~\bibnamefont {Katz}}, \ and\ \bibinfo {author} {\bibfnamefont {J.~L.}\
  \bibnamefont {Shaw}},\ }\href@noop {} {\bibfield  {journal} {\bibinfo
  {journal} {Nat. Photon.}\ }\textbf {\bibinfo {volume} {12}},\ \bibinfo
  {pages} {262} (\bibinfo {year} {2018})}\BibitemShut {NoStop}%
\bibitem [{\citenamefont {Shaltout}\ \emph {et~al.}(2019)\citenamefont
  {Shaltout}, \citenamefont {Lagoudakis}, \citenamefont {{van de G}roep},
  \citenamefont {Shalaev},\ and\ \citenamefont
  {Brongersma}}]{Shaltout19Science}%
  \BibitemOpen
  \bibfield  {author} {\bibinfo {author} {\bibfnamefont {A.~M.}\ \bibnamefont
  {Shaltout}}, \bibinfo {author} {\bibfnamefont {K.~G.}\ \bibnamefont
  {Lagoudakis}}, \bibinfo {author} {\bibfnamefont {J.}~\bibnamefont {{van de
  G}roep}}, \bibinfo {author} {\bibfnamefont {S.~J. K. J. V. V.~M.}\
  \bibnamefont {Shalaev}}, \ and\ \bibinfo {author} {\bibfnamefont {M.~L.}\
  \bibnamefont {Brongersma}},\ }\href@noop {} {\bibfield  {journal} {\bibinfo
  {journal} {Science}\ }\textbf {\bibinfo {volume} {365}},\ \bibinfo {pages}
  {374} (\bibinfo {year} {2019})}\BibitemShut {NoStop}%
\bibitem [{\citenamefont {Hancock}\ \emph {et~al.}(2019)\citenamefont
  {Hancock}, \citenamefont {Zahedpour}, \citenamefont {Goffin},\ and\
  \citenamefont {Milchberg}}]{Hancock19Optica}%
  \BibitemOpen
  \bibfield  {author} {\bibinfo {author} {\bibfnamefont {S.~W.}\ \bibnamefont
  {Hancock}}, \bibinfo {author} {\bibfnamefont {S.}~\bibnamefont {Zahedpour}},
  \bibinfo {author} {\bibfnamefont {A.}~\bibnamefont {Goffin}}, \ and\ \bibinfo
  {author} {\bibfnamefont {H.~M.}\ \bibnamefont {Milchberg}},\ }\href@noop {}
  {\bibfield  {journal} {\bibinfo  {journal} {Optica}\ }\textbf {\bibinfo
  {volume} {6}},\ \bibinfo {pages} {1547} (\bibinfo {year} {2019})}\BibitemShut
  {NoStop}%
\bibitem [{\citenamefont {Jolly}\ \emph {et~al.}(2020)\citenamefont {Jolly},
  \citenamefont {Gobert}, \citenamefont {Jeandet},\ and\ \citenamefont
  {Qu{\'e}r{\'e}}}]{Jolly20OE}%
  \BibitemOpen
  \bibfield  {author} {\bibinfo {author} {\bibfnamefont {S.~W.}\ \bibnamefont
  {Jolly}}, \bibinfo {author} {\bibfnamefont {O.}~\bibnamefont {Gobert}},
  \bibinfo {author} {\bibfnamefont {A.}~\bibnamefont {Jeandet}}, \ and\
  \bibinfo {author} {\bibfnamefont {F.}~\bibnamefont {Qu{\'e}r{\'e}}},\
  }\href@noop {} {\bibfield  {journal} {\bibinfo  {journal} {Opt. Express}\
  }\textbf {\bibinfo {volume} {28}},\ \bibinfo {pages} {4888} (\bibinfo {year}
  {2020})}\BibitemShut {NoStop}%
\bibitem [{\citenamefont {Chong}\ \emph {et~al.}(2020)\citenamefont {Chong},
  \citenamefont {Wan}, \citenamefont {Chen},\ and\ \citenamefont
  {Zhan}}]{Chong20NP}%
  \BibitemOpen
  \bibfield  {author} {\bibinfo {author} {\bibfnamefont {A.}~\bibnamefont
  {Chong}}, \bibinfo {author} {\bibfnamefont {C.}~\bibnamefont {Wan}}, \bibinfo
  {author} {\bibfnamefont {J.}~\bibnamefont {Chen}}, \ and\ \bibinfo {author}
  {\bibfnamefont {Q.}~\bibnamefont {Zhan}},\ }\href@noop {} {\bibfield
  {journal} {\bibinfo  {journal} {Nat. Photon.}\ }\textbf {\bibinfo {volume}
  {14}} (\bibinfo {year} {2020})}\BibitemShut {NoStop}%
\bibitem [{\citenamefont {Chen}\ \emph {et~al.}(2021)\citenamefont {Chen},
  \citenamefont {Wan}, \citenamefont {Chong},\ and\ \citenamefont
  {Zhan}}]{Chen21arxiv}%
  \BibitemOpen
  \bibfield  {author} {\bibinfo {author} {\bibfnamefont {J.}~\bibnamefont
  {Chen}}, \bibinfo {author} {\bibfnamefont {C.}~\bibnamefont {Wan}}, \bibinfo
  {author} {\bibfnamefont {A.}~\bibnamefont {Chong}}, \ and\ \bibinfo {author}
  {\bibfnamefont {Q.}~\bibnamefont {Zhan}},\ }\href@noop {} {\bibfield
  {journal} {\bibinfo  {journal} {arXiv:2101.09452}\ } (\bibinfo {year}
  {2021})}\BibitemShut {NoStop}%
\bibitem [{\citenamefont {Yessenov}\ \emph
  {et~al.}(2019{\natexlab{a}})\citenamefont {Yessenov}, \citenamefont
  {Bhaduri}, \citenamefont {Kondakci},\ and\ \citenamefont
  {Abouraddy}}]{Yessenov19OPN}%
  \BibitemOpen
  \bibfield  {author} {\bibinfo {author} {\bibfnamefont {M.}~\bibnamefont
  {Yessenov}}, \bibinfo {author} {\bibfnamefont {B.}~\bibnamefont {Bhaduri}},
  \bibinfo {author} {\bibfnamefont {H.~E.}\ \bibnamefont {Kondakci}}, \ and\
  \bibinfo {author} {\bibfnamefont {A.~F.}\ \bibnamefont {Abouraddy}},\
  }\href@noop {} {\bibfield  {journal} {\bibinfo  {journal} {Opt. Photon.
  News}\ }\textbf {\bibinfo {volume} {30}},\ \bibinfo {pages} {34} (\bibinfo
  {year} {2019}{\natexlab{a}})}\BibitemShut {NoStop}%
\bibitem [{\citenamefont {Kondakci}\ and\ \citenamefont
  {Abouraddy}(2018{\natexlab{a}})}]{Kondakci18PRL}%
  \BibitemOpen
  \bibfield  {author} {\bibinfo {author} {\bibfnamefont {H.~E.}\ \bibnamefont
  {Kondakci}}\ and\ \bibinfo {author} {\bibfnamefont {A.~F.}\ \bibnamefont
  {Abouraddy}},\ }\href@noop {} {\bibfield  {journal} {\bibinfo  {journal}
  {Phys. Rev. Lett.}\ }\textbf {\bibinfo {volume} {120}},\ \bibinfo {pages}
  {163901} (\bibinfo {year} {2018}{\natexlab{a}})}\BibitemShut {NoStop}%
\bibitem [{\citenamefont {Bhaduri}\ \emph {et~al.}(2018)\citenamefont
  {Bhaduri}, \citenamefont {Yessenov},\ and\ \citenamefont
  {Abouraddy}}]{Bhaduri18OE}%
  \BibitemOpen
  \bibfield  {author} {\bibinfo {author} {\bibfnamefont {B.}~\bibnamefont
  {Bhaduri}}, \bibinfo {author} {\bibfnamefont {M.}~\bibnamefont {Yessenov}}, \
  and\ \bibinfo {author} {\bibfnamefont {A.~F.}\ \bibnamefont {Abouraddy}},\
  }\href@noop {} {\bibfield  {journal} {\bibinfo  {journal} {Opt. Express}\
  }\textbf {\bibinfo {volume} {26}},\ \bibinfo {pages} {20111} (\bibinfo {year}
  {2018})}\BibitemShut {NoStop}%
\bibitem [{\citenamefont {Bhaduri}\ \emph
  {et~al.}(2019{\natexlab{a}})\citenamefont {Bhaduri}, \citenamefont
  {Yessenov}, \citenamefont {Reyes}, \citenamefont {Pena}, \citenamefont
  {Meem}, \citenamefont {Fairchild}, \citenamefont {Menon}, \citenamefont
  {Richardson},\ and\ \citenamefont {Abouraddy}}]{Bhaduri19OL}%
  \BibitemOpen
  \bibfield  {author} {\bibinfo {author} {\bibfnamefont {B.}~\bibnamefont
  {Bhaduri}}, \bibinfo {author} {\bibfnamefont {M.}~\bibnamefont {Yessenov}},
  \bibinfo {author} {\bibfnamefont {D.}~\bibnamefont {Reyes}}, \bibinfo
  {author} {\bibfnamefont {J.}~\bibnamefont {Pena}}, \bibinfo {author}
  {\bibfnamefont {M.}~\bibnamefont {Meem}}, \bibinfo {author} {\bibfnamefont
  {S.~R.}\ \bibnamefont {Fairchild}}, \bibinfo {author} {\bibfnamefont
  {R.}~\bibnamefont {Menon}}, \bibinfo {author} {\bibfnamefont {M.~C.}\
  \bibnamefont {Richardson}}, \ and\ \bibinfo {author} {\bibfnamefont {A.~F.}\
  \bibnamefont {Abouraddy}},\ }\href@noop {} {\bibfield  {journal} {\bibinfo
  {journal} {Opt. Lett.}\ }\textbf {\bibinfo {volume} {44}},\ \bibinfo {pages}
  {2073} (\bibinfo {year} {2019}{\natexlab{a}})}\BibitemShut {NoStop}%
\bibitem [{\citenamefont {Yessenov}\ \emph
  {et~al.}(2019{\natexlab{b}})\citenamefont {Yessenov}, \citenamefont
  {Bhaduri}, \citenamefont {Mach}, \citenamefont {Mardani}, \citenamefont
  {Kondakci}, \citenamefont {Alonso}, \citenamefont {Atia},\ and\ \citenamefont
  {Abouraddy}}]{Yessenov19OE}%
  \BibitemOpen
  \bibfield  {author} {\bibinfo {author} {\bibfnamefont {M.}~\bibnamefont
  {Yessenov}}, \bibinfo {author} {\bibfnamefont {B.}~\bibnamefont {Bhaduri}},
  \bibinfo {author} {\bibfnamefont {L.}~\bibnamefont {Mach}}, \bibinfo {author}
  {\bibfnamefont {D.}~\bibnamefont {Mardani}}, \bibinfo {author} {\bibfnamefont
  {H.~E.}\ \bibnamefont {Kondakci}}, \bibinfo {author} {\bibfnamefont {M.~A.}\
  \bibnamefont {Alonso}}, \bibinfo {author} {\bibfnamefont {G.~A.}\
  \bibnamefont {Atia}}, \ and\ \bibinfo {author} {\bibfnamefont {A.~F.}\
  \bibnamefont {Abouraddy}},\ }\href@noop {} {\bibfield  {journal} {\bibinfo
  {journal} {Opt. Express}\ }\textbf {\bibinfo {volume} {27}},\ \bibinfo
  {pages} {12443} (\bibinfo {year} {2019}{\natexlab{b}})}\BibitemShut {NoStop}%
\bibitem [{\citenamefont {Kondakci}\ and\ \citenamefont
  {Abouraddy}(2019)}]{Kondakci19NC}%
  \BibitemOpen
  \bibfield  {author} {\bibinfo {author} {\bibfnamefont {H.~E.}\ \bibnamefont
  {Kondakci}}\ and\ \bibinfo {author} {\bibfnamefont {A.~F.}\ \bibnamefont
  {Abouraddy}},\ }\href@noop {} {\bibfield  {journal} {\bibinfo  {journal}
  {Nat. Commun.}\ }\textbf {\bibinfo {volume} {10}},\ \bibinfo {pages} {929}
  (\bibinfo {year} {2019})}\BibitemShut {NoStop}%
\bibitem [{\citenamefont {Yessenov}\ \emph
  {et~al.}(2020{\natexlab{a}})\citenamefont {Yessenov}, \citenamefont
  {Bhaduri}, \citenamefont {Delfyett},\ and\ \citenamefont
  {Abouraddy}}]{Yessenov20NC}%
  \BibitemOpen
  \bibfield  {author} {\bibinfo {author} {\bibfnamefont {M.}~\bibnamefont
  {Yessenov}}, \bibinfo {author} {\bibfnamefont {B.}~\bibnamefont {Bhaduri}},
  \bibinfo {author} {\bibfnamefont {P.~J.}\ \bibnamefont {Delfyett}}, \ and\
  \bibinfo {author} {\bibfnamefont {A.~F.}\ \bibnamefont {Abouraddy}},\
  }\href@noop {} {\bibfield  {journal} {\bibinfo  {journal} {Nat. Commun.}\
  }\textbf {\bibinfo {volume} {11}},\ \bibinfo {pages} {5782} (\bibinfo {year}
  {2020}{\natexlab{a}})}\BibitemShut {NoStop}%
\bibitem [{\citenamefont {Bhaduri}\ \emph
  {et~al.}(2019{\natexlab{b}})\citenamefont {Bhaduri}, \citenamefont
  {Yessenov},\ and\ \citenamefont {Abouraddy}}]{Bhaduri19Optica}%
  \BibitemOpen
  \bibfield  {author} {\bibinfo {author} {\bibfnamefont {B.}~\bibnamefont
  {Bhaduri}}, \bibinfo {author} {\bibfnamefont {M.}~\bibnamefont {Yessenov}}, \
  and\ \bibinfo {author} {\bibfnamefont {A.~F.}\ \bibnamefont {Abouraddy}},\
  }\href@noop {} {\bibfield  {journal} {\bibinfo  {journal} {Optica}\ }\textbf
  {\bibinfo {volume} {6}},\ \bibinfo {pages} {139} (\bibinfo {year}
  {2019}{\natexlab{b}})}\BibitemShut {NoStop}%
\bibitem [{\citenamefont {Bhaduri}\ \emph {et~al.}(2020)\citenamefont
  {Bhaduri}, \citenamefont {Yessenov},\ and\ \citenamefont
  {Abouraddy}}]{Bhaduri20NP}%
  \BibitemOpen
  \bibfield  {author} {\bibinfo {author} {\bibfnamefont {B.}~\bibnamefont
  {Bhaduri}}, \bibinfo {author} {\bibfnamefont {M.}~\bibnamefont {Yessenov}}, \
  and\ \bibinfo {author} {\bibfnamefont {A.~F.}\ \bibnamefont {Abouraddy}},\
  }\href@noop {} {\bibfield  {journal} {\bibinfo  {journal} {Nat. Photon.}\
  }\textbf {\bibinfo {volume} {14}},\ \bibinfo {pages} {416} (\bibinfo {year}
  {2020})}\BibitemShut {NoStop}%
\bibitem [{\citenamefont {Shiri}\ \emph
  {et~al.}(2020{\natexlab{a}})\citenamefont {Shiri}, \citenamefont {Yessenov},
  \citenamefont {Webster}, \citenamefont {Schepler},\ and\ \citenamefont
  {Abouraddy}}]{Shiri20NC}%
  \BibitemOpen
  \bibfield  {author} {\bibinfo {author} {\bibfnamefont {A.}~\bibnamefont
  {Shiri}}, \bibinfo {author} {\bibfnamefont {M.}~\bibnamefont {Yessenov}},
  \bibinfo {author} {\bibfnamefont {S.}~\bibnamefont {Webster}}, \bibinfo
  {author} {\bibfnamefont {K.~L.}\ \bibnamefont {Schepler}}, \ and\ \bibinfo
  {author} {\bibfnamefont {A.~F.}\ \bibnamefont {Abouraddy}},\ }\href@noop {}
  {\bibfield  {journal} {\bibinfo  {journal} {Nat. Commun.}\ }\textbf {\bibinfo
  {volume} {11}},\ \bibinfo {pages} {6273} (\bibinfo {year}
  {2020}{\natexlab{a}})}\BibitemShut {NoStop}%
\bibitem [{\citenamefont {Schepler}\ \emph {et~al.}(2020)\citenamefont
  {Schepler}, \citenamefont {Yessenov}, \citenamefont {Zhiyenbayev},\ and\
  \citenamefont {Abouraddy}}]{Schepler20ACSP}%
  \BibitemOpen
  \bibfield  {author} {\bibinfo {author} {\bibfnamefont {K.~L.}\ \bibnamefont
  {Schepler}}, \bibinfo {author} {\bibfnamefont {M.}~\bibnamefont {Yessenov}},
  \bibinfo {author} {\bibfnamefont {Y.}~\bibnamefont {Zhiyenbayev}}, \ and\
  \bibinfo {author} {\bibfnamefont {A.~F.}\ \bibnamefont {Abouraddy}},\
  }\href@noop {} {\bibfield  {journal} {\bibinfo  {journal} {ACS Photon.}\
  }\textbf {\bibinfo {volume} {7}},\ \bibinfo {pages} {2966} (\bibinfo {year}
  {2020})}\BibitemShut {NoStop}%
\bibitem [{\citenamefont {Kondakci}\ and\ \citenamefont
  {Abouraddy}(2018{\natexlab{b}})}]{Kondakci18OL}%
  \BibitemOpen
  \bibfield  {author} {\bibinfo {author} {\bibfnamefont {H.~E.}\ \bibnamefont
  {Kondakci}}\ and\ \bibinfo {author} {\bibfnamefont {A.~F.}\ \bibnamefont
  {Abouraddy}},\ }\href@noop {} {\bibfield  {journal} {\bibinfo  {journal}
  {Opt. Lett.}\ }\textbf {\bibinfo {volume} {43}},\ \bibinfo {pages} {3830}
  (\bibinfo {year} {2018}{\natexlab{b}})}\BibitemShut {NoStop}%
\bibitem [{\citenamefont {Yessenov}\ \emph
  {et~al.}(2020{\natexlab{b}})\citenamefont {Yessenov}, \citenamefont {Hall},
  \citenamefont {Ponomarenko},\ and\ \citenamefont
  {Abouraddy}}]{Yessenov20PRL1}%
  \BibitemOpen
  \bibfield  {author} {\bibinfo {author} {\bibfnamefont {M.}~\bibnamefont
  {Yessenov}}, \bibinfo {author} {\bibfnamefont {L.~A.}\ \bibnamefont {Hall}},
  \bibinfo {author} {\bibfnamefont {S.~A.}\ \bibnamefont {Ponomarenko}}, \ and\
  \bibinfo {author} {\bibfnamefont {A.~F.}\ \bibnamefont {Abouraddy}},\
  }\href@noop {} {\bibfield  {journal} {\bibinfo  {journal} {Phys. Rev. Lett.}\
  }\textbf {\bibinfo {volume} {125}},\ \bibinfo {pages} {243901} (\bibinfo
  {year} {2020}{\natexlab{b}})}\BibitemShut {NoStop}%
\bibitem [{\citenamefont {Yessenov}\ and\ \citenamefont
  {Abouraddy}(2020)}]{Yessenov20PRL2}%
  \BibitemOpen
  \bibfield  {author} {\bibinfo {author} {\bibfnamefont {M.}~\bibnamefont
  {Yessenov}}\ and\ \bibinfo {author} {\bibfnamefont {A.~F.}\ \bibnamefont
  {Abouraddy}},\ }\href@noop {} {\bibfield  {journal} {\bibinfo  {journal}
  {Phys. Rev. Lett.}\ }\textbf {\bibinfo {volume} {125}},\ \bibinfo {pages}
  {233901} (\bibinfo {year} {2020})}\BibitemShut {NoStop}%
\bibitem [{\citenamefont {Yessenov}\ \emph {et~al.}(2021)\citenamefont
  {Yessenov}, \citenamefont {Hall},\ and\ \citenamefont
  {Abouraddy}}]{Yessenov21arxiv}%
  \BibitemOpen
  \bibfield  {author} {\bibinfo {author} {\bibfnamefont {M.}~\bibnamefont
  {Yessenov}}, \bibinfo {author} {\bibfnamefont {L.~A.}\ \bibnamefont {Hall}},
  \ and\ \bibinfo {author} {\bibfnamefont {A.~F.}\ \bibnamefont {Abouraddy}},\
  }\href@noop {} {\bibfield  {journal} {\bibinfo  {journal} {arXiv:2102.09443}\
  } (\bibinfo {year} {2021})}\BibitemShut {NoStop}%
\bibitem [{\citenamefont {Yessenov}\ \emph
  {et~al.}(2019{\natexlab{c}})\citenamefont {Yessenov}, \citenamefont
  {Bhaduri}, \citenamefont {Kondakci},\ and\ \citenamefont
  {Abouraddy}}]{Yessenov19PRA}%
  \BibitemOpen
  \bibfield  {author} {\bibinfo {author} {\bibfnamefont {M.}~\bibnamefont
  {Yessenov}}, \bibinfo {author} {\bibfnamefont {B.}~\bibnamefont {Bhaduri}},
  \bibinfo {author} {\bibfnamefont {H.~E.}\ \bibnamefont {Kondakci}}, \ and\
  \bibinfo {author} {\bibfnamefont {A.~F.}\ \bibnamefont {Abouraddy}},\
  }\href@noop {} {\bibfield  {journal} {\bibinfo  {journal} {Phys. Rev. A}\
  }\textbf {\bibinfo {volume} {99}},\ \bibinfo {pages} {023856} (\bibinfo
  {year} {2019}{\natexlab{c}})}\BibitemShut {NoStop}%
\bibitem [{\citenamefont {Shaarawi}\ \emph {et~al.}(2000)\citenamefont
  {Shaarawi}, \citenamefont {Besieris}, \citenamefont {Attiya},\ and\
  \citenamefont {El-Diwany}}]{Shaarawi00JASA}%
  \BibitemOpen
  \bibfield  {author} {\bibinfo {author} {\bibfnamefont {A.~M.}\ \bibnamefont
  {Shaarawi}}, \bibinfo {author} {\bibfnamefont {I.~M.}\ \bibnamefont
  {Besieris}}, \bibinfo {author} {\bibfnamefont {A.~M.}\ \bibnamefont
  {Attiya}}, \ and\ \bibinfo {author} {\bibfnamefont {E.}~\bibnamefont
  {El-Diwany}},\ }\href@noop {} {\bibfield  {journal} {\bibinfo  {journal} {J.
  Acoust. Soc. Am.}\ }\textbf {\bibinfo {volume} {107}},\ \bibinfo {pages} {70}
  (\bibinfo {year} {2000})}\BibitemShut {NoStop}%
\bibitem [{\citenamefont {Attiya}\ \emph {et~al.}(2001)\citenamefont {Attiya},
  \citenamefont {El-Diwany}, \citenamefont {Shaarawi},\ and\ \citenamefont
  {Besieris}}]{Attiya01PER}%
  \BibitemOpen
  \bibfield  {author} {\bibinfo {author} {\bibfnamefont {A.~M.}\ \bibnamefont
  {Attiya}}, \bibinfo {author} {\bibfnamefont {E.}~\bibnamefont {El-Diwany}},
  \bibinfo {author} {\bibfnamefont {A.~M.}\ \bibnamefont {Shaarawi}}, \ and\
  \bibinfo {author} {\bibfnamefont {I.~M.}\ \bibnamefont {Besieris}},\
  }\href@noop {} {\bibfield  {journal} {\bibinfo  {journal} {Prog. Electromagn.
  Res.}\ }\textbf {\bibinfo {volume} {30}},\ \bibinfo {pages} {191} (\bibinfo
  {year} {2001})}\BibitemShut {NoStop}%
\bibitem [{\citenamefont {Shaarawi}\ \emph {et~al.}(2001)\citenamefont
  {Shaarawi}, \citenamefont {Besieris}, \citenamefont {Attiya},\ and\
  \citenamefont {El-Diwany}}]{Shaarawi01PER}%
  \BibitemOpen
  \bibfield  {author} {\bibinfo {author} {\bibfnamefont {A.~M.}\ \bibnamefont
  {Shaarawi}}, \bibinfo {author} {\bibfnamefont {I.~M.}\ \bibnamefont
  {Besieris}}, \bibinfo {author} {\bibfnamefont {A.~M.}\ \bibnamefont
  {Attiya}}, \ and\ \bibinfo {author} {\bibfnamefont {E.}~\bibnamefont
  {El-Diwany}},\ }\href@noop {} {\bibfield  {journal} {\bibinfo  {journal}
  {Prog. Electromagn. Res.}\ }\textbf {\bibinfo {volume} {30}},\ \bibinfo
  {pages} {213} (\bibinfo {year} {2001})}\BibitemShut {NoStop}%
\bibitem [{\citenamefont {Salem}\ and\ \citenamefont {Ba{\v
  g}c{\i}}(2012)}]{Salem12JOSAA}%
  \BibitemOpen
  \bibfield  {author} {\bibinfo {author} {\bibfnamefont {M.~A.}\ \bibnamefont
  {Salem}}\ and\ \bibinfo {author} {\bibfnamefont {H.}~\bibnamefont {Ba{\v
  g}c{\i}}},\ }\href@noop {} {\bibfield  {journal} {\bibinfo  {journal} {J.
  Opt. Soc. Am. A}\ }\textbf {\bibinfo {volume} {29}},\ \bibinfo {pages} {139}
  (\bibinfo {year} {2012})}\BibitemShut {NoStop}%
\bibitem [{\citenamefont {Hillion}(1993)}]{Hillion93Optik}%
  \BibitemOpen
  \bibfield  {author} {\bibinfo {author} {\bibfnamefont {P.}~\bibnamefont
  {Hillion}},\ }\href@noop {} {\bibfield  {journal} {\bibinfo  {journal}
  {Optik}\ }\textbf {\bibinfo {volume} {93}},\ \bibinfo {pages} {67} (\bibinfo
  {year} {1993})}\BibitemShut {NoStop}%
\bibitem [{\citenamefont {Donnelly}\ and\ \citenamefont
  {Power}(1997)}]{Donnelly97IEEE}%
  \BibitemOpen
  \bibfield  {author} {\bibinfo {author} {\bibfnamefont {R.}~\bibnamefont
  {Donnelly}}\ and\ \bibinfo {author} {\bibfnamefont {D.}~\bibnamefont
  {Power}},\ }\href@noop {} {\bibfield  {journal} {\bibinfo  {journal} {IEEE
  Trans. Antennas Propag.}\ }\textbf {\bibinfo {volume} {45}},\ \bibinfo
  {pages} {580} (\bibinfo {year} {1997})}\BibitemShut {NoStop}%
\bibitem [{\citenamefont {Hillion}(1998)}]{Hillion98JO}%
  \BibitemOpen
  \bibfield  {author} {\bibinfo {author} {\bibfnamefont {P.}~\bibnamefont
  {Hillion}},\ }\href@noop {} {\bibfield  {journal} {\bibinfo  {journal} {J.
  Opt.}\ }\textbf {\bibinfo {volume} {29}},\ \bibinfo {pages} {345} (\bibinfo
  {year} {1998})}\BibitemShut {NoStop}%
\bibitem [{\citenamefont {Hillion}(1999)}]{Hillion99JOA}%
  \BibitemOpen
  \bibfield  {author} {\bibinfo {author} {\bibfnamefont {P.}~\bibnamefont
  {Hillion}},\ }\href@noop {} {\bibfield  {journal} {\bibinfo  {journal} {J.
  Opt. A}\ }\textbf {\bibinfo {volume} {1}},\ \bibinfo {pages} {459} (\bibinfo
  {year} {1999})}\BibitemShut {NoStop}%
\bibitem [{\citenamefont {Saleh}\ and\ \citenamefont
  {Teich}(2019)}]{SalehBook07}%
  \BibitemOpen
  \bibfield  {author} {\bibinfo {author} {\bibfnamefont {B.~E.~A.}\
  \bibnamefont {Saleh}}\ and\ \bibinfo {author} {\bibfnamefont {M.~C.}\
  \bibnamefont {Teich}},\ }\href@noop {} {\emph {\bibinfo {title} {Fundamentals
  of Photonics}}}\ (\bibinfo  {publisher} {Wiley},\ \bibinfo {year}
  {2019})\BibitemShut {NoStop}%
\bibitem [{\citenamefont {Torres}\ \emph {et~al.}(2010)\citenamefont {Torres},
  \citenamefont {Hendrych},\ and\ \citenamefont {Valencia}}]{Torres10AOP}%
  \BibitemOpen
  \bibfield  {author} {\bibinfo {author} {\bibfnamefont {J.~P.}\ \bibnamefont
  {Torres}}, \bibinfo {author} {\bibfnamefont {M.}~\bibnamefont {Hendrych}}, \
  and\ \bibinfo {author} {\bibfnamefont {A.}~\bibnamefont {Valencia}},\
  }\href@noop {} {\bibfield  {journal} {\bibinfo  {journal} {Adv. Opt.
  Photon.}\ }\textbf {\bibinfo {volume} {2}},\ \bibinfo {pages} {319} (\bibinfo
  {year} {2010})}\BibitemShut {NoStop}%
\bibitem [{\citenamefont {F{\"u}l{\"o}p}\ and\ \citenamefont
  {Hebling}(2010)}]{Fulop10Review}%
  \BibitemOpen
  \bibfield  {author} {\bibinfo {author} {\bibfnamefont {J.~A.}\ \bibnamefont
  {F{\"u}l{\"o}p}}\ and\ \bibinfo {author} {\bibfnamefont {J.}~\bibnamefont
  {Hebling}},\ }in\ \href@noop {} {\emph {\bibinfo {booktitle} {Recent Optical
  and Photonic Technologies}}},\ \bibinfo {editor} {edited by\ \bibinfo
  {editor} {\bibfnamefont {K.~Y.}\ \bibnamefont {Kim}}}\ (\bibinfo  {publisher}
  {InTech},\ \bibinfo {year} {2010})\BibitemShut {NoStop}%
\bibitem [{\citenamefont {Reivelt}\ and\ \citenamefont
  {Saari}(2000)}]{Reivelt00JOSAA}%
  \BibitemOpen
  \bibfield  {author} {\bibinfo {author} {\bibfnamefont {K.}~\bibnamefont
  {Reivelt}}\ and\ \bibinfo {author} {\bibfnamefont {P.}~\bibnamefont
  {Saari}},\ }\href@noop {} {\bibfield  {journal} {\bibinfo  {journal} {J. Opt.
  Soc. Am. A}\ }\textbf {\bibinfo {volume} {17}},\ \bibinfo {pages} {1785}
  (\bibinfo {year} {2000})}\BibitemShut {NoStop}%
\bibitem [{\citenamefont {Reivelt}\ and\ \citenamefont
  {Saari}(2002)}]{Reivelt02PRE}%
  \BibitemOpen
  \bibfield  {author} {\bibinfo {author} {\bibfnamefont {K.}~\bibnamefont
  {Reivelt}}\ and\ \bibinfo {author} {\bibfnamefont {P.}~\bibnamefont
  {Saari}},\ }\href@noop {} {\bibfield  {journal} {\bibinfo  {journal} {Phys.
  Rev. E}\ }\textbf {\bibinfo {volume} {66}},\ \bibinfo {pages} {056611}
  (\bibinfo {year} {2002})}\BibitemShut {NoStop}%
\bibitem [{\citenamefont {Shaarawi}\ and\ \citenamefont
  {Besieris}(2000)}]{Shaarawi00JPA}%
  \BibitemOpen
  \bibfield  {author} {\bibinfo {author} {\bibfnamefont {A.~M.}\ \bibnamefont
  {Shaarawi}}\ and\ \bibinfo {author} {\bibfnamefont {I.~M.}\ \bibnamefont
  {Besieris}},\ }\href@noop {} {\bibfield  {journal} {\bibinfo  {journal} {J.
  Phys. A}\ }\textbf {\bibinfo {volume} {33}},\ \bibinfo {pages} {7255}
  (\bibinfo {year} {2000})}\BibitemShut {NoStop}%
\bibitem [{\citenamefont {Saari}(2018)}]{SaariPRA18}%
  \BibitemOpen
  \bibfield  {author} {\bibinfo {author} {\bibfnamefont {P.}~\bibnamefont
  {Saari}},\ }\href@noop {} {\bibfield  {journal} {\bibinfo  {journal} {Phys.
  Rev. A}\ }\textbf {\bibinfo {volume} {97}},\ \bibinfo {pages} {063824}
  (\bibinfo {year} {2018})}\BibitemShut {NoStop}%
\bibitem [{\citenamefont {Saari}\ \emph {et~al.}(2019)\citenamefont {Saari},
  \citenamefont {Rebane},\ and\ \citenamefont {Besieris}}]{Saari19PRA}%
  \BibitemOpen
  \bibfield  {author} {\bibinfo {author} {\bibfnamefont {P.}~\bibnamefont
  {Saari}}, \bibinfo {author} {\bibfnamefont {O.}~\bibnamefont {Rebane}}, \
  and\ \bibinfo {author} {\bibfnamefont {I.}~\bibnamefont {Besieris}},\
  }\href@noop {} {\bibfield  {journal} {\bibinfo  {journal} {Phys. Rev. A}\
  }\textbf {\bibinfo {volume} {100}},\ \bibinfo {pages} {013849} (\bibinfo
  {year} {2019})}\BibitemShut {NoStop}%
\bibitem [{\citenamefont {Saari}\ and\ \citenamefont
  {Besieris}(2020)}]{Saari20PRA}%
  \BibitemOpen
  \bibfield  {author} {\bibinfo {author} {\bibfnamefont {P.}~\bibnamefont
  {Saari}}\ and\ \bibinfo {author} {\bibfnamefont {I.}~\bibnamefont
  {Besieris}},\ }\href@noop {} {\bibfield  {journal} {\bibinfo  {journal}
  {Phys. Rev. A}\ }\textbf {\bibinfo {volume} {101}},\ \bibinfo {pages}
  {023812} (\bibinfo {year} {2020})}\BibitemShut {NoStop}%
\bibitem [{\citenamefont {Besieris}\ \emph {et~al.}(1998)\citenamefont
  {Besieris}, \citenamefont {Abdel-Rahman}, \citenamefont {Shaarawi},\ and\
  \citenamefont {Chatzipetros}}]{Besieris98PIERS}%
  \BibitemOpen
  \bibfield  {author} {\bibinfo {author} {\bibfnamefont {I.}~\bibnamefont
  {Besieris}}, \bibinfo {author} {\bibfnamefont {M.}~\bibnamefont
  {Abdel-Rahman}}, \bibinfo {author} {\bibfnamefont {A.}~\bibnamefont
  {Shaarawi}}, \ and\ \bibinfo {author} {\bibfnamefont {A.}~\bibnamefont
  {Chatzipetros}},\ }\href@noop {} {\bibfield  {journal} {\bibinfo  {journal}
  {Progr. in Electrom. Res.}\ }\textbf {\bibinfo {volume} {19}},\ \bibinfo
  {pages} {1} (\bibinfo {year} {1998})}\BibitemShut {NoStop}%
\bibitem [{\citenamefont {B{\'e}langer}(1984)}]{Belanger84JOSAA}%
  \BibitemOpen
  \bibfield  {author} {\bibinfo {author} {\bibfnamefont {P.~A.}\ \bibnamefont
  {B{\'e}langer}},\ }\href@noop {} {\bibfield  {journal} {\bibinfo  {journal}
  {J. Opt. Soc. Am. A}\ }\textbf {\bibinfo {volume} {1}},\ \bibinfo {pages}
  {723} (\bibinfo {year} {1984})}\BibitemShut {NoStop}%
\bibitem [{\citenamefont {Bowlan}\ \emph {et~al.}(2009)\citenamefont {Bowlan},
  \citenamefont {Valtna-Lukner}, \citenamefont {L{\~o}hmus}, \citenamefont
  {Piksarv}, \citenamefont {Saari},\ and\ \citenamefont
  {Trebino}}]{Bowlan09OL}%
  \BibitemOpen
  \bibfield  {author} {\bibinfo {author} {\bibfnamefont {P.}~\bibnamefont
  {Bowlan}}, \bibinfo {author} {\bibfnamefont {H.}~\bibnamefont
  {Valtna-Lukner}}, \bibinfo {author} {\bibfnamefont {M.}~\bibnamefont
  {L{\~o}hmus}}, \bibinfo {author} {\bibfnamefont {P.}~\bibnamefont {Piksarv}},
  \bibinfo {author} {\bibfnamefont {P.}~\bibnamefont {Saari}}, \ and\ \bibinfo
  {author} {\bibfnamefont {R.}~\bibnamefont {Trebino}},\ }\href@noop {}
  {\bibfield  {journal} {\bibinfo  {journal} {Opt. Lett.}\ }\textbf {\bibinfo
  {volume} {34}},\ \bibinfo {pages} {2276} (\bibinfo {year}
  {2009})}\BibitemShut {NoStop}%
\bibitem [{\citenamefont {Bonaretti}\ \emph {et~al.}(2009)\citenamefont
  {Bonaretti}, \citenamefont {Faccio}, \citenamefont {Clerici}, \citenamefont
  {Biegert},\ and\ \citenamefont {{Di T}rapani}}]{Bonaretti09OE}%
  \BibitemOpen
  \bibfield  {author} {\bibinfo {author} {\bibfnamefont {F.}~\bibnamefont
  {Bonaretti}}, \bibinfo {author} {\bibfnamefont {D.}~\bibnamefont {Faccio}},
  \bibinfo {author} {\bibfnamefont {M.}~\bibnamefont {Clerici}}, \bibinfo
  {author} {\bibfnamefont {J.}~\bibnamefont {Biegert}}, \ and\ \bibinfo
  {author} {\bibfnamefont {P.}~\bibnamefont {{Di T}rapani}},\ }\href@noop {}
  {\bibfield  {journal} {\bibinfo  {journal} {Opt. Express}\ }\textbf {\bibinfo
  {volume} {17}},\ \bibinfo {pages} {9804} (\bibinfo {year}
  {2009})}\BibitemShut {NoStop}%
\bibitem [{\citenamefont {Kuntz}\ \emph {et~al.}(2009)\citenamefont {Kuntz},
  \citenamefont {Braverman}, \citenamefont {Youn}, \citenamefont {Lobino},
  \citenamefont {Pessina},\ and\ \citenamefont {Lvovsky}}]{Kuntz09PRA}%
  \BibitemOpen
  \bibfield  {author} {\bibinfo {author} {\bibfnamefont {K.~B.}\ \bibnamefont
  {Kuntz}}, \bibinfo {author} {\bibfnamefont {B.}~\bibnamefont {Braverman}},
  \bibinfo {author} {\bibfnamefont {S.~H.}\ \bibnamefont {Youn}}, \bibinfo
  {author} {\bibfnamefont {M.}~\bibnamefont {Lobino}}, \bibinfo {author}
  {\bibfnamefont {E.~M.}\ \bibnamefont {Pessina}}, \ and\ \bibinfo {author}
  {\bibfnamefont {A.~I.}\ \bibnamefont {Lvovsky}},\ }\href@noop {} {\bibfield
  {journal} {\bibinfo  {journal} {Phys. Rev. A}\ }\textbf {\bibinfo {volume}
  {79}},\ \bibinfo {pages} {043802} (\bibinfo {year} {2009})}\BibitemShut
  {NoStop}%
\bibitem [{\citenamefont {{Allende M}otz}\ \emph {et~al.}(2021)\citenamefont
  {{Allende M}otz}, \citenamefont {Yessenov},\ and\ \citenamefont
  {Abouraddy}}]{Motz21arxiv}%
  \BibitemOpen
  \bibfield  {author} {\bibinfo {author} {\bibfnamefont {A.~M.}\ \bibnamefont
  {{Allende M}otz}}, \bibinfo {author} {\bibfnamefont {M.}~\bibnamefont
  {Yessenov}}, \ and\ \bibinfo {author} {\bibfnamefont {A.~F.}\ \bibnamefont
  {Abouraddy}},\ }\href@noop {} {\bibfield  {journal} {\bibinfo  {journal}
  {arXiv:2102.10505}\ } (\bibinfo {year} {2021})}\BibitemShut {NoStop}%
\bibitem [{\citenamefont {Zamboni-Rached}\ \emph {et~al.}(2001)\citenamefont
  {Zamboni-Rached}, \citenamefont {Recami},\ and\ \citenamefont
  {Fontana}}]{Zamboni01PRE}%
  \BibitemOpen
  \bibfield  {author} {\bibinfo {author} {\bibfnamefont {M.}~\bibnamefont
  {Zamboni-Rached}}, \bibinfo {author} {\bibfnamefont {E.}~\bibnamefont
  {Recami}}, \ and\ \bibinfo {author} {\bibfnamefont {F.}~\bibnamefont
  {Fontana}},\ }\href@noop {} {\bibfield  {journal} {\bibinfo  {journal} {Phys.
  Rev E}\ }\textbf {\bibinfo {volume} {64}},\ \bibinfo {pages} {066603}
  (\bibinfo {year} {2001})}\BibitemShut {NoStop}%
\bibitem [{\citenamefont {Zamboni-Rached}\ \emph {et~al.}(2003)\citenamefont
  {Zamboni-Rached}, \citenamefont {Fontana},\ and\ \citenamefont
  {Recami}}]{Zamboni03PRE}%
  \BibitemOpen
  \bibfield  {author} {\bibinfo {author} {\bibfnamefont {M.}~\bibnamefont
  {Zamboni-Rached}}, \bibinfo {author} {\bibfnamefont {F.}~\bibnamefont
  {Fontana}}, \ and\ \bibinfo {author} {\bibfnamefont {E.}~\bibnamefont
  {Recami}},\ }\href@noop {} {\bibfield  {journal} {\bibinfo  {journal} {Phys.
  Rev. E}\ }\textbf {\bibinfo {volume} {67}},\ \bibinfo {pages} {036620}
  (\bibinfo {year} {2003})}\BibitemShut {NoStop}%
\bibitem [{\citenamefont {Shabahang}\ \emph {et~al.}(2017)\citenamefont
  {Shabahang}, \citenamefont {Kondakci}, \citenamefont {Villinger},
  \citenamefont {Perlstein}, \citenamefont {{El H}alawany},\ and\ \citenamefont
  {Abouraddy}}]{Shabahang17SR}%
  \BibitemOpen
  \bibfield  {author} {\bibinfo {author} {\bibfnamefont {S.}~\bibnamefont
  {Shabahang}}, \bibinfo {author} {\bibfnamefont {H.~E.}\ \bibnamefont
  {Kondakci}}, \bibinfo {author} {\bibfnamefont {M.~L.}\ \bibnamefont
  {Villinger}}, \bibinfo {author} {\bibfnamefont {J.~D.}\ \bibnamefont
  {Perlstein}}, \bibinfo {author} {\bibfnamefont {A.}~\bibnamefont {{El
  H}alawany}}, \ and\ \bibinfo {author} {\bibfnamefont {A.~F.}\ \bibnamefont
  {Abouraddy}},\ }\href@noop {} {\bibfield  {journal} {\bibinfo  {journal}
  {Sci. Rep.}\ }\textbf {\bibinfo {volume} {7}},\ \bibinfo {pages} {10336}
  (\bibinfo {year} {2017})}\BibitemShut {NoStop}%
\bibitem [{\citenamefont {Villinger}\ \emph {et~al.}(2019)\citenamefont
  {Villinger}, \citenamefont {Shiri}, \citenamefont {Shabahang}, \citenamefont
  {Jahromi}, \citenamefont {Nasr}, \citenamefont {Villinger},\ and\
  \citenamefont {Abouraddy}}]{Villinger19unpubl}%
  \BibitemOpen
  \bibfield  {author} {\bibinfo {author} {\bibfnamefont {M.~L.}\ \bibnamefont
  {Villinger}}, \bibinfo {author} {\bibfnamefont {A.}~\bibnamefont {Shiri}},
  \bibinfo {author} {\bibfnamefont {S.}~\bibnamefont {Shabahang}}, \bibinfo
  {author} {\bibfnamefont {A.~K.}\ \bibnamefont {Jahromi}}, \bibinfo {author}
  {\bibfnamefont {M.~B.}\ \bibnamefont {Nasr}}, \bibinfo {author}
  {\bibfnamefont {C.}~\bibnamefont {Villinger}}, \ and\ \bibinfo {author}
  {\bibfnamefont {A.~F.}\ \bibnamefont {Abouraddy}},\ }\href@noop {} {\bibfield
   {journal} {\bibinfo  {journal} {arXiv:1911.09276}\ } (\bibinfo {year}
  {2019})}\BibitemShut {NoStop}%
\bibitem [{\citenamefont {Shiri}\ \emph
  {et~al.}(2020{\natexlab{b}})\citenamefont {Shiri}, \citenamefont {Yessenov},
  \citenamefont {Aravindakshan},\ and\ \citenamefont {Abouraddy}}]{Shiri20OL}%
  \BibitemOpen
  \bibfield  {author} {\bibinfo {author} {\bibfnamefont {A.}~\bibnamefont
  {Shiri}}, \bibinfo {author} {\bibfnamefont {M.}~\bibnamefont {Yessenov}},
  \bibinfo {author} {\bibfnamefont {R.}~\bibnamefont {Aravindakshan}}, \ and\
  \bibinfo {author} {\bibfnamefont {A.~F.}\ \bibnamefont {Abouraddy}},\
  }\href@noop {} {\bibfield  {journal} {\bibinfo  {journal} {Opt. Lett.}\
  }\textbf {\bibinfo {volume} {45}},\ \bibinfo {pages} {1774} (\bibinfo {year}
  {2020}{\natexlab{b}})}\BibitemShut {NoStop}%
\bibitem [{\citenamefont {Shiri}\ \emph
  {et~al.}(2020{\natexlab{c}})\citenamefont {Shiri}, \citenamefont {Schepler},\
  and\ \citenamefont {Abouraddy}}]{Shiri20APLP}%
  \BibitemOpen
  \bibfield  {author} {\bibinfo {author} {\bibfnamefont {A.}~\bibnamefont
  {Shiri}}, \bibinfo {author} {\bibfnamefont {K.~L.}\ \bibnamefont {Schepler}},
  \ and\ \bibinfo {author} {\bibfnamefont {A.~F.}\ \bibnamefont {Abouraddy}},\
  }\href@noop {} {\bibfield  {journal} {\bibinfo  {journal} {APL Photon.}\
  }\textbf {\bibinfo {volume} {5}},\ \bibinfo {pages} {106107} (\bibinfo {year}
  {2020}{\natexlab{c}})}\BibitemShut {NoStop}%
\end{thebibliography}%

\end{document}